# Stable stochastic dynamics in yeast cell cycle

Yurie Okabe* and Masaki Sasai*§

*Department of Computational Science and Engineering, Nagoya University, Nagoya 464-8603, Japan

§Core Research for Evolutional Science and Technology-Japan Science and Technology Agency, Nagoya 464-8603, Japan

**ABSTRACT**

Chemical reactions in cell are subject to intense stochastic fluctuations. An important question is how the fundamental physiological behavior of cell is kept stable against those noisy perturbations. In this paper a stochastic model of cell cycle of budding yeast is constructed to analyze the effects of noise on the cell cycle oscillation. The model predicts intense noise in levels of mRNAs and proteins, and the simulated protein levels explain the observed statistical tendency of noise in populations of synchronous and asynchronous cells. In spite of intense noise in levels of proteins and mRNAs, cell cycle is stable enough to bring the largely perturbed cells back to the physiological cyclic oscillation. The model shows that consecutively appearing fixed points are the origin of this stability of cell cycle.

## INTRODUCTION

Noisy fluctuations are inevitable features of chemical reactions in cell, which should lead to cell-to-cell variation in a genetically identical population of cells (1-3). One of the important issues in modern cell biology is to understand how the molecular reaction network bearing such noisy fluctuations produces the orchestrated behavior for functioning. In this paper we take cell cycle of budding yeast, *Saccharomyces cerevisiae*, as an example to analyze how its dynamics tolerates noise to maintain a coherent cyclic oscillation.

The cell cycle mechanism is well conserved among eukaryotes (4), where the cyclic ups and downs of activity of complexes of cyclins and cyclin-dependent kinases (CDKs) are at the heart of its dynamics (5). The reaction network regulating the cyclin/CDK activity, however, includes many positive and negative feedback loops, which is too complex to be verbalized, so that the mathematical modeling of the reaction network is necessary (6). Tyson and colleagues have constructed models of cell cycle of budding yeast (7, 8), fission yeast (9, 10), and frog eggs (11) by describing networks of reaction kinetics with differential equations. Their model of budding yeast describes cell cycle as transitions between two stable states (7, 8) as has been hypothesized by Nasmyth (12). Li *et al*., on the other hand, described cell cycle of budding yeast with a network of Boolean functions (13). In this model the cell-cycle dynamics is represented by trajectories of the Boolean states, which shift toward a fixed point corresponding to the biologically stable G1 phase. Although these deterministic models have clarified important aspects (14), effects of stochasticity still largely remain to be resolved.

Noise tolerance of a checkpoint mechanism in cell cycle has been discussed theoretically (15) and robustness of stochastic models of cell cycle of budding yeast (16) and fission yeast (17) has been studied. In these models, however, noise has been introduced as a given disturbance of the deterministic kinetic rules and the mechanism to generate the noise has not been discussed. In the present work, noise is described as a dynamical feature that is inevitable in the model and the strength of noise that should occur in cell cycle is estimated to clarify the mechanism which ensures the stability against thus generated noise.

Fluctuations in protein numbers in budding yeast have been measured by decomposing fluctuations into intrinsic and extrinsic noises (1, 18, 19), where intrinsic noise has been defined as fluctuations which arise from smallness of numbers of molecules in reactions. Extrinsic noise has been the rest part originating from the fluctuating physiological condition (20). In this paper we consider both intrinsic and extrinsic noises by regarding the intrinsic noise as fluctuations arising from the stochastic dynamics of reactions in the regulation network of biomolecules and the extrinsic noise as those arising from the mechanisms working outside of the network. In prokaryote, combination of intrinsic and extrinsic noises in simulation has given a quantitative explanation of the experimentally observed protein levels (21). We use a similar approach although processes involved here are much more complex.

Our goal in the present paper is to clarify the mechanism of noise tolerance of the cyclic oscillation by using thus developed stochastic model of cell cycle.

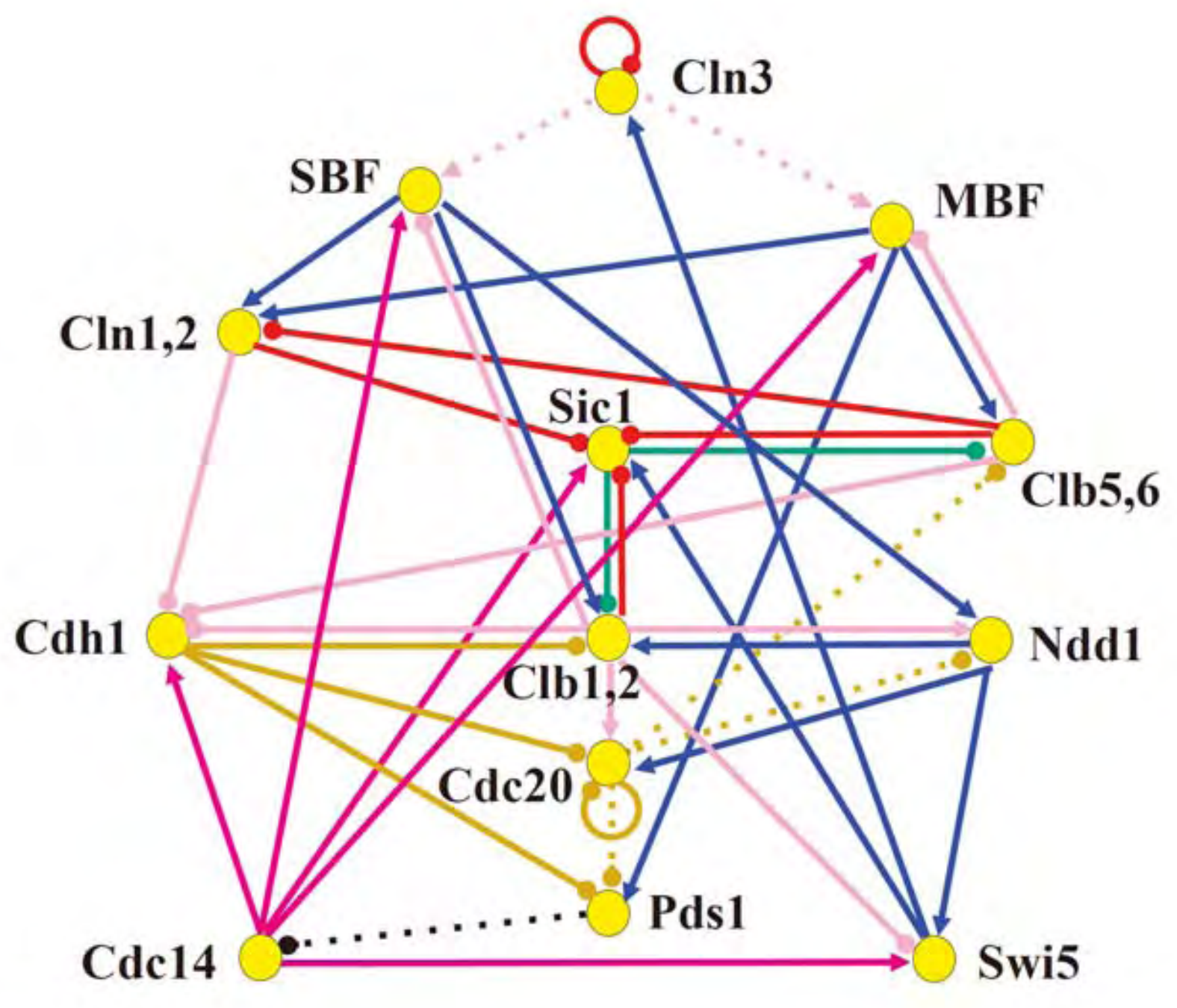

**Figure 1.** A model of the reaction network which sustains cell cycle of budding yeast. Each node represents a gene and its product, mRNA and protein. Arrows with a triangular head denote positive regulations, whereas arrows with a round head show negative regulations. Colors of arrows specify the types of regulations: transcriptional regulation (blue), phosphorylation (pink), dephosphorylation (dark pink), ubiquitination (light brown), phosphorylation as a mark of ubiquitination (red), protein-complex formation (green), and suppression of diffusion (black). Cdc28, which is CDK in budding yeast, is abundant through cell cycle and hence is not explicitly considered in the model. Cln1 and Cln2 are assumed to work in combination and hence treated as a unit, Cln1,2, in the model. Clb1,2 and Clb5,6 are also treated as units, respectively. The dotted arrows are assumed to work only in specific stages: phosphorylation of SBF and MBF by Cln3 (stage1), ubiquitination of Clb5,6, Ndd1, and Pds1 triggered by Cdc20 (stages3, 4, 5), and suppression of diffusion of Cdc14 by Pds1 (stage4).

**STOCHASTIC MODEL OF CELL CYCLE**

In order to address the questions of noise in cell cycle, the budding yeast cell cycle is modeled as shown in Fig.1, where each node represents a gene and its products, *i.e.* mRNA and protein. Transcription and translation are modeled at each node by the stochastic kinetic processes. Each link is the transcriptional regulation or the post-transcriptional regulation such as phosphorylation, dephosphorylation, ubiquitination, or complex formation. The network includes 13 proteins which have been considered in Ref.13. Although the whole biomolecular network relevant to cell cycle is gigantic including more than 800 relevant genes (22), here only the essential part of it is abstracted. Marginal interactions between the network components in the model and those in other reactions in cell are treated as constraints imposed on the model. See *Supporting Text*1 for the catalog of molecular species and reactions in the model. There are still many important details in transcriptional and translational processes which are not explicitly considered in the model, such as chromatin remodeling or nucleosome replacement. The simplified coarse-grained modeling to neglect these aspects, however, was successful in quantitatively describing dynamics of small regulatory networks in yeast cell (18, 19), and we may expect that the similar coarse-graining provides insights on the present complex network as well.

Intrinsic noise is treated by describing the network state with three types of variables; states of genes, numbers of mRNA molecules, and numbers of protein molecules. We write $\xi(\mu) = 1$ or "the $\mu$th gene is on" when the transcription factors are bound to the promoter of the $\mu$th gene, and $\xi(\mu) = 0$ or "the $\mu$th gene is off", otherwise. Transcription rates of 11 genes of Fig.1, $\mu$ = *PDS1*, *CLN1,2*, *CLN3*, *CLB1,2*, *CLB5,6*, *SIC1*, *CDC20*, *SWI5*, and *NDD1*, are controlled by transcriptional factors in the network, so that each of them is transcribed with a high rate when $\xi(\mu) = 1$ and with a low rate when $\xi(\mu) = 0$. Other four genes are assumed to be transcribed constitutively with a mild transcription rate: $\xi(\mu)$ is fixed to be $\xi(\mu) = 1$ for $\mu$ = *CDH1*, *CDC14*, *MBF*, and *SBF*. See *Supplementary Table*1 for the values of the transcription rate constant. The state of the $\mu$th gene, $\alpha(\mu)$, is defined as $\alpha(\mu) = \xi(\mu)$ before the $\mu$th gene is duplicated, and $\alpha(\mu) = (\xi(\mu)\ \xi'(\mu)) = (1,1), (1,0), (0,1)$ or $(0,0)$ after the $\mu$th gene is duplicated.

The master equation is derived for the probability distribution of states of genes, numbers of mRNA molecules, and numbers of protein molecules residing in each of chemical states. Equations for their moments are derived and those equations are treated approximately by truncating them at the 2nd order of cumulants and by neglecting the cross correlation between different molecular species. See *Supporting Text*2 for the concrete form of the equations. The network dynamics is then numerically followed by solving a set of differential equations for means and variance; the mean number of mRNA molecules transcribed from the $\mu$th gene of the state $\alpha$ at time $t$, $N_{m\alpha}^{\mathrm{int}}(\mu, t)$, variance of the number of mRNA molecules, $\sigma_{m\alpha}^{\mathrm{int}}(\mu, t)^2$, the mean number of $\mu$th protein molecules at the chemical state $X$, $N_X^{\mathrm{int}}(\mu, t)$, variance of the number of protein molecules, $\sigma_X^{\mathrm{int}}(\mu, t)^2$, and probability that the $\mu$th gene is at state $\alpha$, $D_\alpha^{\mathrm{int}}(\mu, t)$. Here, the suffix "int" stresses that averages are taken over the fluctuations caused by intrinsic noise. $X$ denotes the chemical state of whether the protein is phosphorylated, dephosphorelated, or ubiquitinated. See Appendix for the precise definition of chemical states. Differential equations for means and variances are numerically solved to estimate the effects of the intrinsic noise. Factors such as $F_{m\alpha}^{\mathrm{nt}}(\mu, t) = \sigma_{m\alpha}^{\mathrm{int}}(\mu, t)^2/N_{m\alpha}^{\mathrm{int}}(\mu, t)$ and $F_X^{\mathrm{int}}(\mu, t) = \sigma_X^{\mathrm{int}}(\mu, t)^2/N_X^{\mathrm{int}}(\mu, t)$ measure the strength of intrinsic noise.

A benchmark test of the truncated cumulant approximation introduced above is carried out by taking a small reaction network in Fig.2 as an example system. The truncated cumulant approximation is applied to this system and the results are compared in Fig.3 with the exact numerical simulation of the corresponding master equation. The truncated cumulant approximation agrees well with the numerical simulation for $N_{m\alpha}^{\mathrm{int}}(\mu, t)$, $\sigma_{m\alpha}^{\mathrm{int}}(\mu, t)^2$, $N_X^{\mathrm{int}}(\mu, t)$, and $D_\alpha^{\mathrm{int}}(\mu, t)$, but the approximation tends to underestimate $\sigma_X^{\mathrm{int}}(\mu, t)^2$. In spite of such systematic deviation, we can find in Fig.3 that the approximation used here gives reasonable estimation for both $F_{m\alpha}^{\mathrm{int}}(\mu, t)$ and $F_X^{\mathrm{int}}(\mu, t)$.

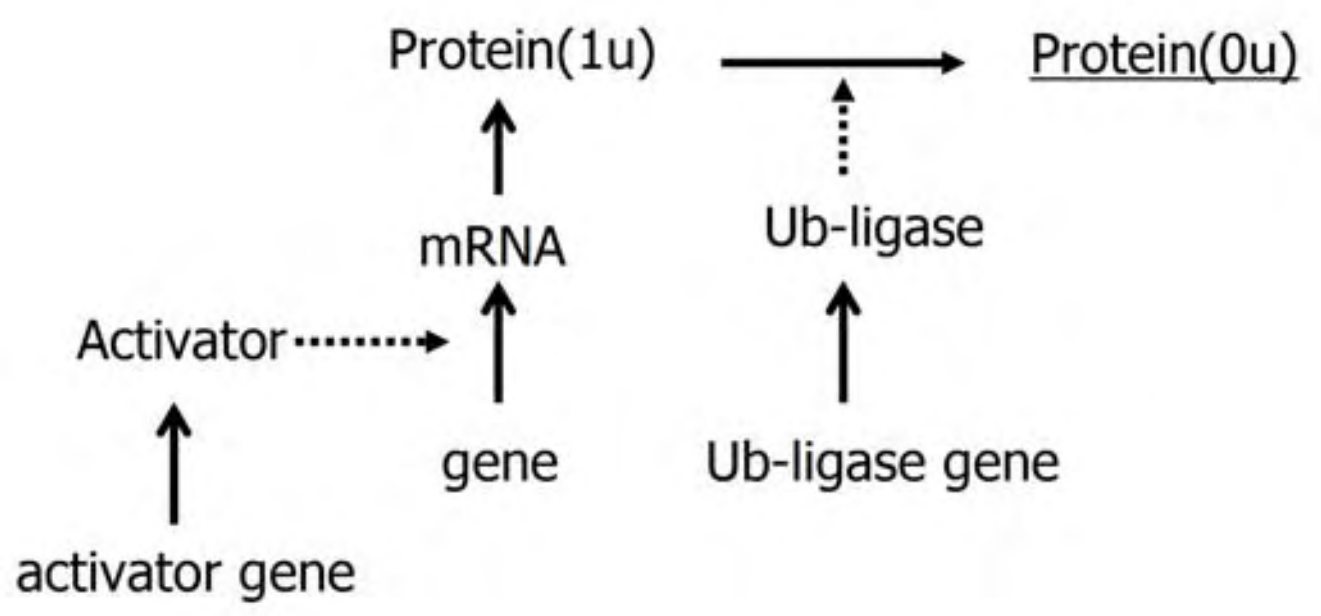

**Figure 2.** A reaction system to test the truncated cumulant approximation. The synthesis rate of activator and that of ubiquitin ligase are modulated by $\sin(2\pi t/T)$ to mimic the cell cycle oscillation with a typical period of $T$ =125 min. When activator is bound to the promoter of the gene, the gene is turned on to synthesize mRNA, which then yields Protein(1u). When Protein(1u) is ubiquitinated through the act of ubiquitin ligase, Protein(1u) is turned into Protein(0u). The unstable short-lived protein is underlined. Although mRNA and all proteins are assumed to be degraded with certain specific rates in the model, those degradation processes are omitted from this figure. Coefficients of reaction rates are same as in *Supplementary Table*1 except for the temporally modulated synthesis rates of activator and ubiquitin ligase.

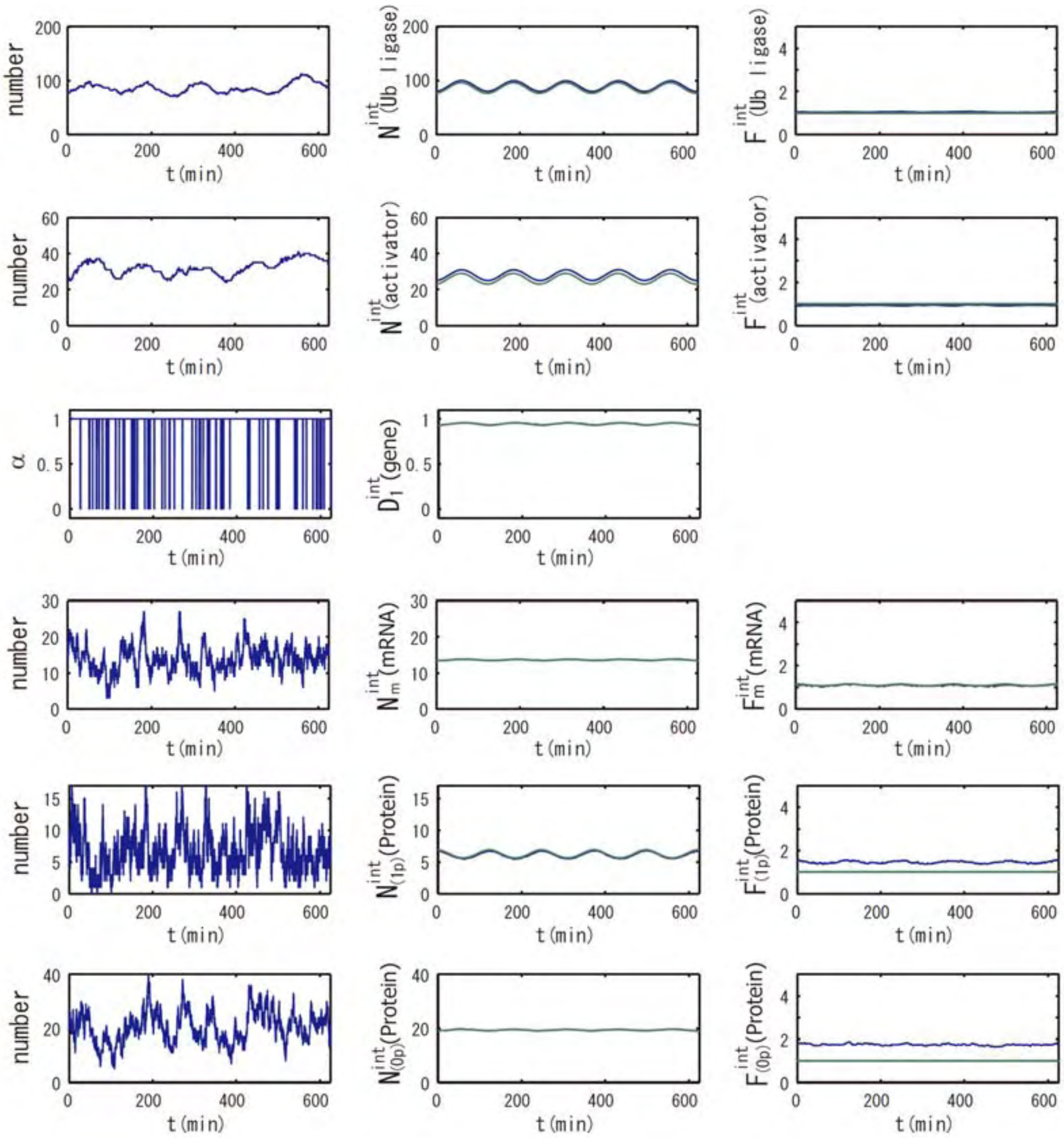

**Figure 3.** Comparison of the truncated cumulant approximation and the numerical Monte Carlo (MC) simulation. The MC simulation was performed by employing the Gillespie algorithm (31) to numerically solve the master equation which describes the reaction processes of Fig.2. (Left column) An example of trajectory of the numerical MC simulation. From top to bottom, the number of ubiquitin ligase, the number of activator, the number of mRNA, the number of Protein(1u), and the number of Protein(0u) are shown as functions of time. (Middle column) The mean number of corresponding molecules obtained by averaging $10^4$ MC trajectories (blue lines) are compared with the mean number of molecules obtained by using the truncated cumulant approximation (green lines). (Right column) The Fano factor *i.e.*, ratio of variance to mean of the number of molecules obtained by sampling $10^4$ MC trajectories (blue lines) is compared with that obtained by using the truncated cumulant approximation (green lines). From top to bottom, the Fano factors of ubiquitin ligase, activator, Protein(1u), and Protein(0u) are shown as functions of time.

As sources of the extrinsic noise, we consider several types of events; regulations at checkpoints, release of Cdc14 at the late anaphase, DNA replication, and cell division. During cell cycle, these events occur in stochastic manners, which perturb and diversify the trajectories of $\{N_{m\alpha}^{\text{int}}(\mu, t), \sigma_{m\alpha}^{\text{int}}(\mu, t)^2, N_X^{\text{int}}(\mu, t), \sigma_X^{\text{int}}(\mu, t)^2$, and $D_\alpha^{\text{int}}(\mu, t)\}$. Strength of extrinsic noise is estimated from diversity of trajectories of $\{N_{m\alpha}^{\text{int}}(\mu,t)\}$ and $\{N_X^{\text{int}}(\mu,t)\}$ as $\sigma_{m\alpha}^{\text{ext}}(\mu,t)^2 = \langle N_{m\alpha}^{\text{int}}(\mu,t)^2\rangle - \langle N_{m\alpha}^{\text{int}}(\mu,t)\rangle^2$ and $\sigma_X^{\text{ext}}(\mu,t)^2 = \langle N_X^{\text{int}}(\mu,t)^2\rangle - \langle N_X^{\text{int}}(\mu,t)\rangle^2$, where $\langle\ldots\rangle$ is average over an ensemble of trajectories. Then, the total cell-to-cell variances are $\sigma_{m\alpha}^{\text{total}}(\mu, t)^2 = \langle\sigma_{m\alpha}^{\text{int}}(\mu, t)^2\rangle + \sigma_{m\alpha}^{\text{ext}}(\mu, t)^2$ and $\sigma_X^{\text{total}}(\mu, t)^2 = \langle\sigma_X^{\text{int}}(\mu, t)^2\rangle + \sigma_X^{\text{ext}}(\mu, t)^2$.

In cell cycle the checkpoint mechanisms bridge between reactions in the network and physiological changes in cell. For example, the spindle-assembly checkpoint blocks onset of anaphase by suppressing the activity of Cdc20 in the network until properly attached chromosomes have lined up on the metaphase plate in the center of the spindle (23). We here consider checkpoints to monitor the following events or conditions: ($C_1$) sufficient cell growth to start DNA replication, ($C_2$) completion of DNA replication, and ($C_3$) spindle assembly. In addition to these checkpoints, mitotic exit is tightly controlled by the release of Cdc14 from nucleolus and the protein numbers are drastically changed by cell division. We refer to the release of Cdc14 as $C_4$ and cell division as $C_5$. We refer to the duration between $C_i$ and $C_{i+1}$ ($i$ = 1-4) as stage$i$ and the duration between $C_5$ and $C_1$ as stage5. See Fig.4 for the definition of stages. Although there can be other cellular-level events or conditions whose details have not yet been clarified, we treat $C_1$-$C_5$ as representative examples to see how these events perturb the network dynamics. In the present model, effects of the cellular-level events are expressed by modulations of reactions: Some reactions are allowed only before or after passing certain $C_i$, or in other words, the network in Fig.1 has some specific links which are validated only for certain specific stages. Duration of stage$i$ in the $r$th round of cell cycle, $T_r(i)$, is determined as a random number fluctuating in the range $0.8 \le T_r(i)/T_0(i) \le 1.2$, where $T_0(i)$ is the standard value of duration inferred from experiments; $T_0(1)$=40min, $T_0(2)$=15min, $T_0(3)$=20min, $T_0(4)$=10min, and $T_0(5)$=40min (24-26). In this way the structure of the differential equations is modulated when the system passes through {$C_i$} at the fluctuating timing. The fluctuation in timing works as extrinsic noise posed to the network.

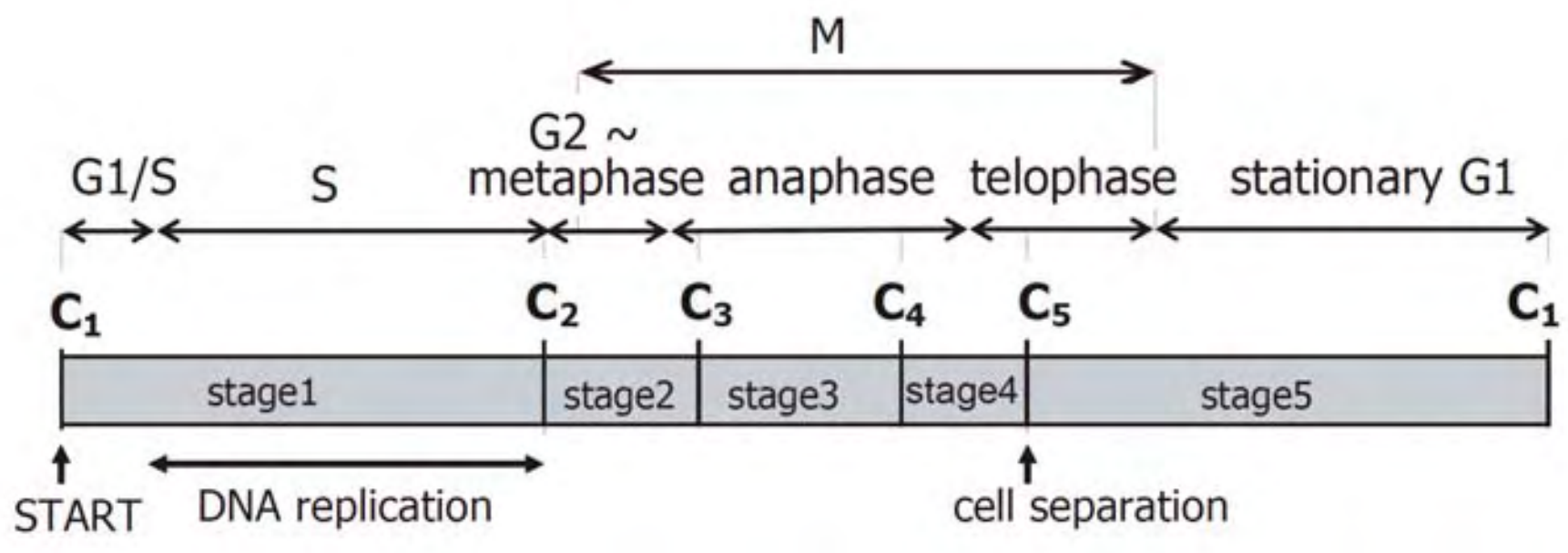

**Figure 4.** Definition of stages. Stages delimited by five cellular events $C_{1\text{-}5}$ are compared with the cell-cycle phases of usual terminology. In budding yeast, boundary between S and G2 or that between G2 and M is vague.

DNA replication and cell division are other sources of extrinsic noise. In stage1 DNA is replicated and each of 13 genes in the network is doubled. The time when each gene is duplicated is randomly selected at each round of cell cycle between the time 10 minutes past $C_1$ and the end of stage1. After DNA is replicated, budding yeast cells undergo far less chromosomal condensation than animal cells and the nuclear envelope remains intact throughout the cell cycle, so that the transcription rate is kept high even in mitosis (27). After passing $C_5$ the duplicated DNA and other molecules are distributed to daughter and mother cells. Although there is a temporal gap of several minutes between the nuclear separation and cytokinesis in real cells (25, 26), we do not distinguish their timing for simplicity. In the simulation, duplicated 13 genes are equally distributed to daughter and mother but the volume ratio between separated nuclei should bear fluctuations to some extent (25, 26). We assume that the ratio is randomly fluctuating in the range from 1:1 to 0.9:1.1. Proteins which are localized in nucleus are handed to the daughter according to this ratio. Cytokinesis should be fluctuating with a larger amplitude than the nuclear separation, so that we assume that mRNAs and proteins which may locate in cytoplasm are distributed to daughter and mother with a fluctuating ratio from 1:1 to 0.6:1.4.

In this way both intrinsic and extrinsic noises are dynamically generated in the model. In the following, the statistical features of thus generated noises are compared with experiments to investigate how cell cycle maintains the stable oscillation under the influence of these noises.

The network model of Fig.1 includes more than 300 rate constants of reactions. Although we may be able to fit the individual experimental data by calibrating these parameters, such detailed comparison with experiments is not the purpose of the present paper. Our goal here is to quantify the statistical tendency of intrinsic and extrinsic noises to analyze the basic mechanism to ensure the persistency of cyclic dynamics. In order to focus on such mechanism, we adopt a simplified parameterization by categorizing reactions into 15 different types and assigning a single parameter to each type. These reactions and parameters are explained in *Supplementary Table*1.

## RESULTS

### Cell-cycle attractor

The five cellular events ($C_{1\text{-}5}$) were chosen as the initial starting points of the simulation. For each initial time point, 1000 initial values were randomly generated in the ranges, $0\leq D_\alpha^{\mathrm{int}}(\mu,0)\leq 1$, $0\leq N_{m\alpha}^{\mathrm{int}}(\mu,0)\leq 20N_{\mathrm{g}}$, $0\leq \sigma_{m\alpha}^{\mathrm{int}}(\mu,0)^2\leq 20N_{m\alpha}^{\mathrm{int}}(\mu,0)$, $0\leq N_X^{\mathrm{int}}(\mu,0)\leq 100$, and $0\leq \sigma_X^{\mathrm{int}}(\mu,0)^2\leq 10N_X^{\mathrm{int}}(\mu,0)$, where $N_{\mathrm{g}}$ is the number of copies of genes in a cell; $N_{\mathrm{g}} = 1$ for $C_1$ and $C_5$, and $N_{\mathrm{g}} = 2$ for $C_2$, $C_3$ and $C_4$. From all of tested 5000 initial conditions, the simulated trajectories converged to a narrow region in the solution space and showed an oscillatory motion. In this narrow region the numbers of mRNA molecules, $\Sigma_\alpha D_\alpha^{\mathrm{int}}(\mu, t)N_{m\alpha}^{\mathrm{int}}(\mu,t)$, were roughly in the range from 0 to 30 and most of the numbers of protein molecules at the chemical state $X$, $N_X^{\mathrm{int}}(\mu,t)$, were in the range from 0 to 75, leading to the accumulated oscillation of $\Sigma_X N_X^{\mathrm{int}}(\mu,t)$ from 0 to 130. We refer to this attractive region in the solution space as the cell-cycle attractor. Examples of 5 trajectories starting at $C_1$ are shown in Fig.5a by projecting them onto the space of three mean numbers of proteins. With this representation, the cell-cycle attractor appears as a doughnut-shaped region in the three-dimensional space.

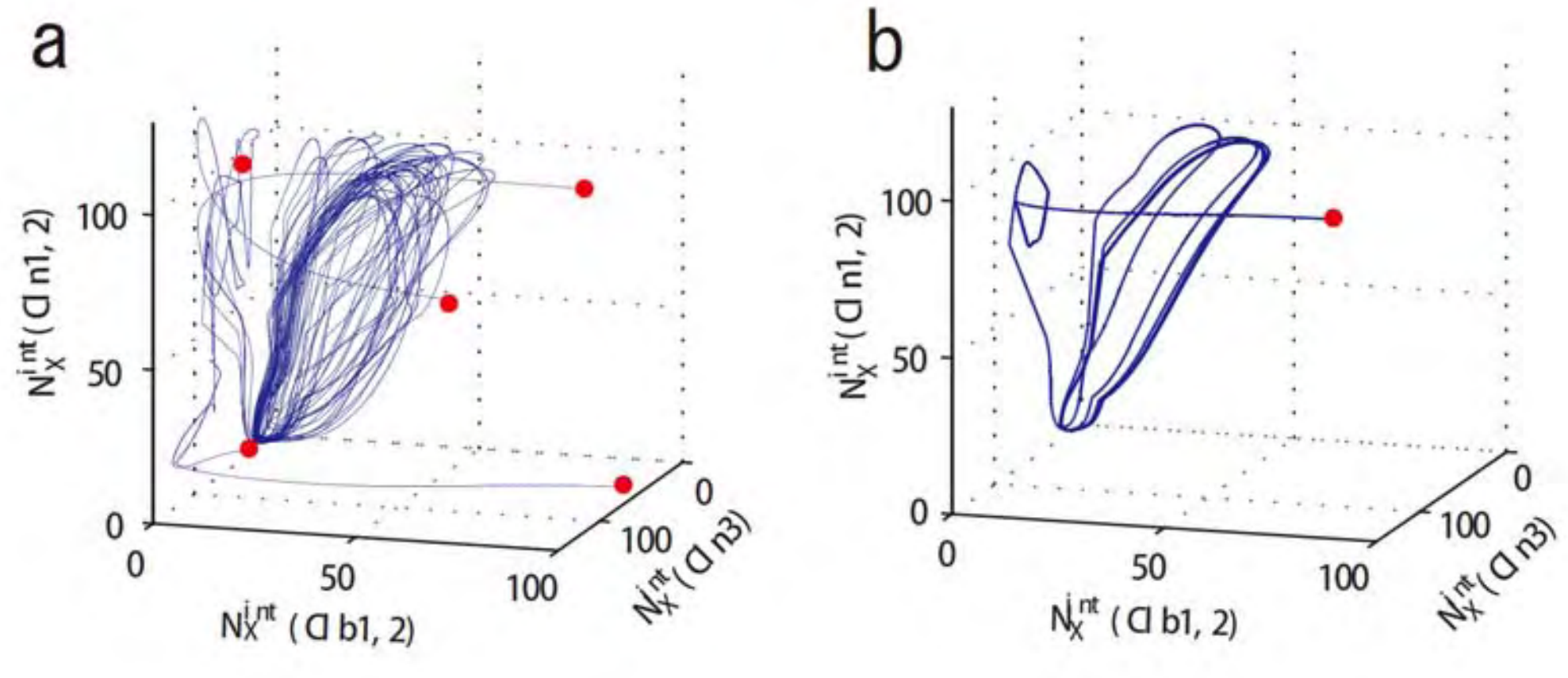

**Figure 5**. Convergence of trajectories to the cell-cycle attractor. Trajectories are projected onto the three dimensional space of $N_X^{\mathrm{int}}$(Cln3, $t$) with $X$=(0p)(1u), $N_X^{\mathrm{int}}$(Clb1,2, $t$) with $X$=(1u), and $N_X^{\mathrm{int}}$(Cln1,2, $t$) with $X$=(0p)(1u). See Appendix for the definition of $X$. (**a**) Five trajectories starting at $C_1$ with random initial conditions (red stars) are attracted to the cell-cycle attractor. (**b**) Under the constraint that extrinsic noise is absent, trajectories converge to the cell-cycle attractor to form a limit cycle.

The convergent behavior of trajectories suggests that a stable closed orbit of the cyclic oscillation is hidden behind the cell-cycle attractor, which becomes clear when the external noise is turned off with the constraints; (i) Durations of stages are fixed to the standard values. (ii) The 13 genes are duplicated at the fixed timing with the fixed order. (iii) The volume ratio of nuclei and that of cytoplasm in cell separation are both fixed to be 1:1. Under these constraints trajectories converge to a closed orbit with $\sigma_{m\alpha}^{\mathrm{ext}}(\mu, t)^2 = \sigma_X^{\mathrm{ext}}(\mu, t)^2 = 0$ as shown in Fig.5b. We call this orbit the standard limit cycle. This standard limit cycle underlies the cell-cycle attractor around which trajectories are attracted under the influence of extrinsic noise.

Robustness of the standard limit cycle was tested by changing parameters one by one from the standard values. The limit cycle remains stable when those parameters are between MIN and MAX shown in *Supplementary Table*1. For many parameters of post-translational reactions, the ratio MAX/MIN exceeds $10^3$. This robustness should partly justify our rough estimation of 15 grouped parameters instead of the precise determination of many individual parameters. For parameters relevant to the transcription and translation processes, this ratio is around 2-3, indicating the importance of rather strict transcriptional regulations to maintain cell cycle.

Stochastic trajectories attracted to the cell-cycle attractor are consistent with the observed cell cycle oscillation. In Fig.6 the mean numbers of three proteins, Clb2, Clb5, and Sic1, calculated under the influence of extrinsic noise are shown. Here, the transcription rate of Clb5 in the model is adjusted to be smaller than transcription rates of other proteins by a factor of 0.5 to obtain the apparent agreement between the simulated peak height of the Clb5 number and the observed data (14). Other features such as the small amount of Sic1 and the timing that each protein number shows a peak do not depend on this calibration. See *Supplementary Tables* 2 and 3 to compare the simulated and observed data for other mRNAs and proteins.

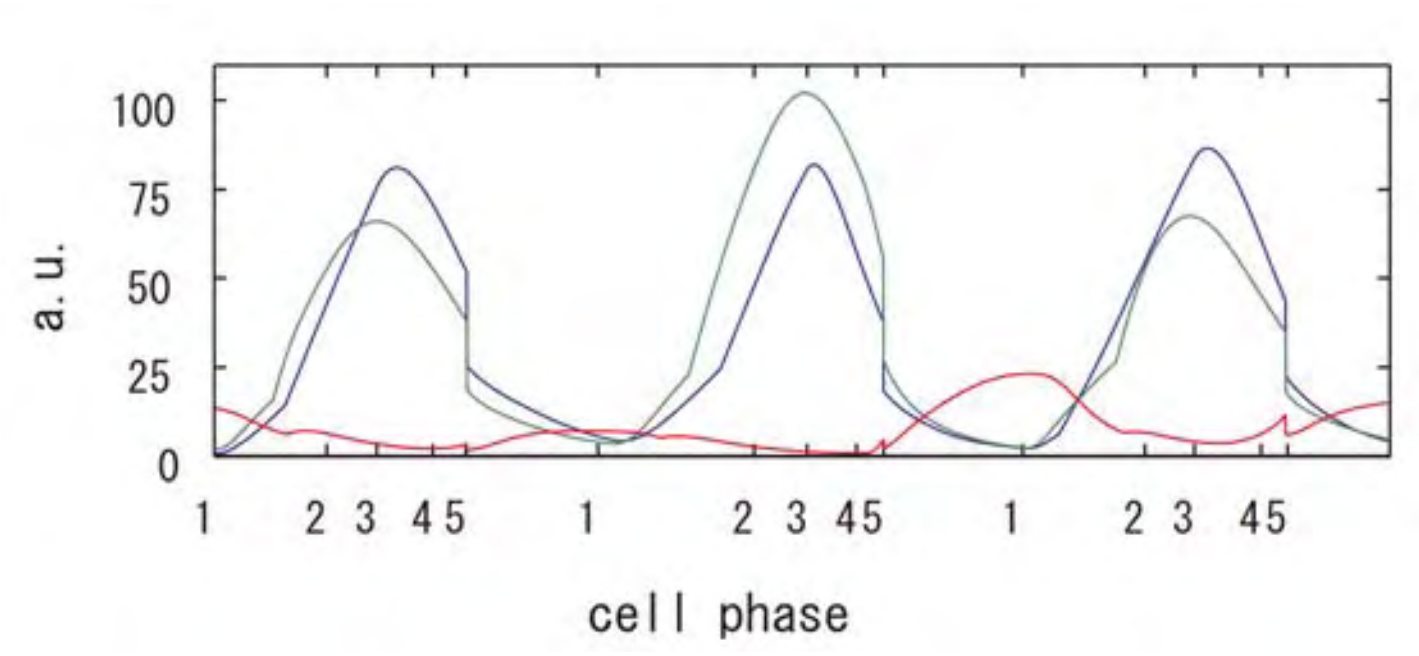

**Figure 6.** Temporal change of the mean numbers of Clb5,6, Clb1,2, and Sic1. blue: $\Sigma_X N_X^{\mathrm{int}}(\mathrm{Clb5,6},\ t)+\Sigma_X N_X^{\mathrm{int}}(\mathrm{Clb5,6/Sic1},\ t)$, green: $\Sigma_X N_X^{\mathrm{int}}(\mathrm{Clb1,2},\ t)+\Sigma_X N_X^{\mathrm{int}}(\mathrm{Clb1,2/Sic1},\ t)$, and red: $\Sigma_X N_X^{\mathrm{int}}(\mathrm{Sic1},\ t)+\Sigma_X N_X^{\mathrm{int}}(\mathrm{Clb5,6/Sic1},\ t)+ \Sigma_X N_X^{\mathrm{int}}(\mathrm{Clb1,2/Sic1},\ t)$. $i$=1-5 on the horizontal axis indicates cellular events $C_i$.

**Intrinsic and extrinsic noises**

Strength of intrinsic and extrinsic noises can be quantified from the simulated results, which should then provide a basis to understand the stability of the cell-cycle attractor against these noises.

Strength of intrinsic noise was measured by $F_{m\alpha}^{\mathrm{int}}(\mu, t)$ and $F_X^{\mathrm{int}}(\mu, t)$ calculated along the simulated trajectories. $F_{m\alpha}^{\mathrm{int}}(\mu, t)$ oscillates with the amplitude of $0< F_{m\alpha}^{\mathrm{int}}(\mu, t) <10$, for $\mu$ = *CLN3*, *SIC1*, *CLN12*, *CLB56*, *PDS1*, and *CLB12* and with the amplitude of $4<F_{m\alpha}^{\mathrm{int}}(\mu, t)<10$ for $\mu$= *SWI5* and *CDC20*. $F_X^{\mathrm{int}}(\mu, t)$ for proteins involved in autocatalytic reactions, $\mu$ = Cln3 and Cdc20, oscillates with $0< F_X^{\mathrm{int}}(\mu, t) <10$. For other 11 proteins, $F_X^{\mathrm{int}}(\mu, t)$ rapidly converges to unity and remains almost constant throughout the cell cycle. Although $F_X^{\mathrm{int}}(\mu, t)$ tends to be underestimated in the present approximation, we should stress that $F_X^{\mathrm{int}}(\mu, t)$ for the latter 11 proteins is kept smaller than that for Cln3 and Cdc20. Such modest $F_X^{\mathrm{int}}(\mu, t)$ for many proteins implies that the design of the network which does not contain many autocatalytic loops or the small-length positive feed-back loops effectively reduces intrinsic noise to prevent $F_X^{\mathrm{int}}(\mu, t)$ from being too large. In this way the intrinsic noise in protein levels is suppressed, which stabilizes the cell-cycle attractor. Intrinsic noise in RNA levels is larger than that in protein levels giving wider distributions than Poissonian. Such difference between $F_{m\alpha}^{\mathrm{int}}(\mu, t)$ and $F_X^{\mathrm{int}}(\mu, t)$ is consistent with the frequently observed difference between transcriptome and proteome (28).

Strength of extrinsic noise, $F_X^{\mathrm{ext}}(\mu, t)=\sigma_X^{\mathrm{ext}}(\mu, t)^2/N_X(\mu,t)$, can be estimated by sampling trajectories fluctuating around the standard limit cycle. Here, $N_X(\mu,t) = \langle N_X^{\mathrm{int}}(\mu,t)\rangle$, and $\langle\ldots\rangle$ is average over an ensemble of trajectories. Temporal change of $F_X^{\mathrm{ext}}(\mu, t)$ is shown in Fig.7a for an ensemble of trajectories starting from $C_1$ at $t = 0$. Although the individual $F_X^{\mathrm{ext}}(\mu, t)$ depends on $\mu$ and $X$ in characteristic ways, extrinsic noise accumulates as time proceeds, which randomly shifts the phase of each trajectory to increase $F_X^{\mathrm{ext}}(\mu, t)$. In the large $t$ limit, trajectories are completely dephased to make $F_X^{\mathrm{ext}}(\mu, t)$ constant as shown in Fig.7b. This effect is more evident when the average is taken over $\mu$ and $X$ as shown in Figs.7c and d. Thus, the extrinsic noise is small when cells are synchronous having similar phases and largest when cells are completely dephased. This difference between the ensemble of synchronous cells and that of asynchronous cells is shown in Figs.8a and 8c by plotting histograms of $F_X^{\mathrm{ext}}(\mu)$ for those ensembles. Also shown are histograms of $\sigma_X^{\mathrm{ext}}(\mu)^2/\sigma_X^{\mathrm{int}}(\mu)^2$ averaged over ensembles of synchronous (Fig.8b) and asynchronous (Fig.8d) cells. Fig.8 indicates that intrinsic noise is important when synchronous cells are sampled and extrinsic noise dominates when asynchronous cells are sampled.

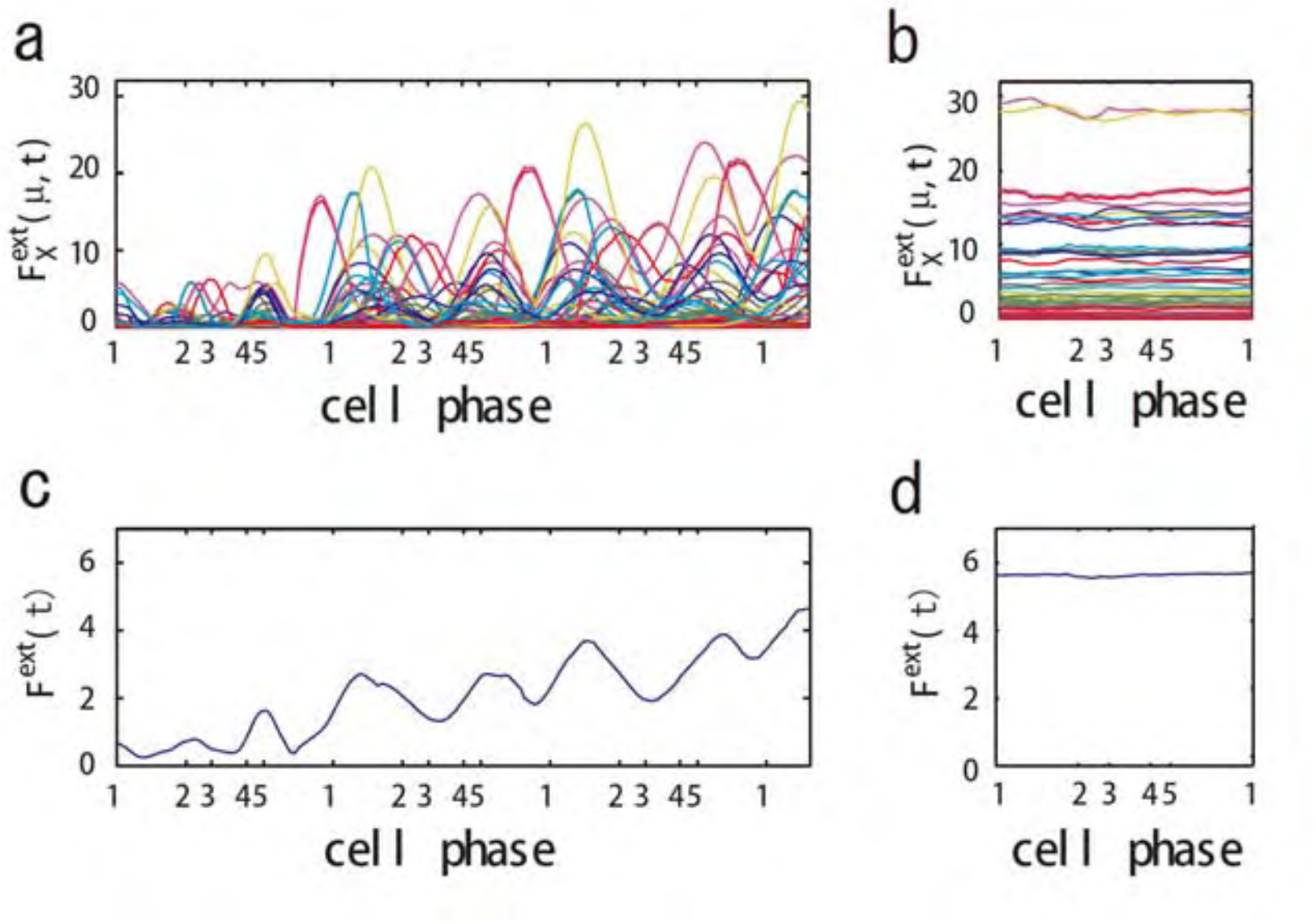

**Figure 7.** Dephasing and increase of extrinsic noise. $F_X^{\text{ext}}(\mu, t)$ is averaged over 1200 trajectories starting at the same cell-cycle phase. Extrinsic noise accumulates over time due to the dephasing of trajectories (**a** and **c**). In the large $t$ limit, trajectories are completely dephased to make $F_X^{\text{ext}}(\mu, t)$ almost constant (**b** and **d**). In **c** and **d**, $F_X^{\text{ext}}(\mu, t)$ are averaged over $\mu$ and $X$. $i$=1-5 on the horizontal axis indicates the average time of passing $C_i$.

Such dominance of intrinsic or extrinsic noise can be verified by comparing the calculated results with the experimental data. In Ref.1 a proteome-wide measurement of fluctuations of protein levels were reported by sorting cells according to their size. The sorting was performed by gating the cell flow to select cells smaller than the gate size. Since the cell size is smallest just after cell division and increases through cell cycle, gated cells should correspond to cells just after $C_5$ in simulation. Averages over ungated cells should be the averages over asynchronous cells. In Fig.9, the simulated results of $CV(\mu, X)^2 = \sigma_X^{\text{total}}(\mu)^2/N_X(\mu)^2$ are plotted as functions of $N_X(\mu)$ for both gated and ungated cases, where $\sigma_X^{\text{total}}(\mu)^2 = \langle\sigma_X^{\text{int}}(\mu)^2\rangle + \sigma_X^{\text{ext}}(\mu)^2$. The extrinsic noise is reduced by gating and the feature of constant $F_X^{\text{int}}(\mu, t)$ is manifested in the plot to make $CV(\mu, X)^2$ roughly proportional to $1/N_X(\mu)$. The same feature of $CV^2 \approx 1/N$ was observed in the gated data of Ref.1. We should note, however, that $\sigma_X^{\text{ext}}(\mu)^2$ does not completely vanish even when the cell phase is specified as in gated cells, which is consistent with the observation in Refs.1 and 19. In Fig.9 CV$(\mu, X)^2$ for ungated cells is dominated by the extrinsic noise and takes values around $10^{3.5}$ with weaker dependence on $N_X(\mu)$ as was observed in Ref.1. Thus, the present model quantitatively reproduces observed features of intrinsic and extrinsic noises.

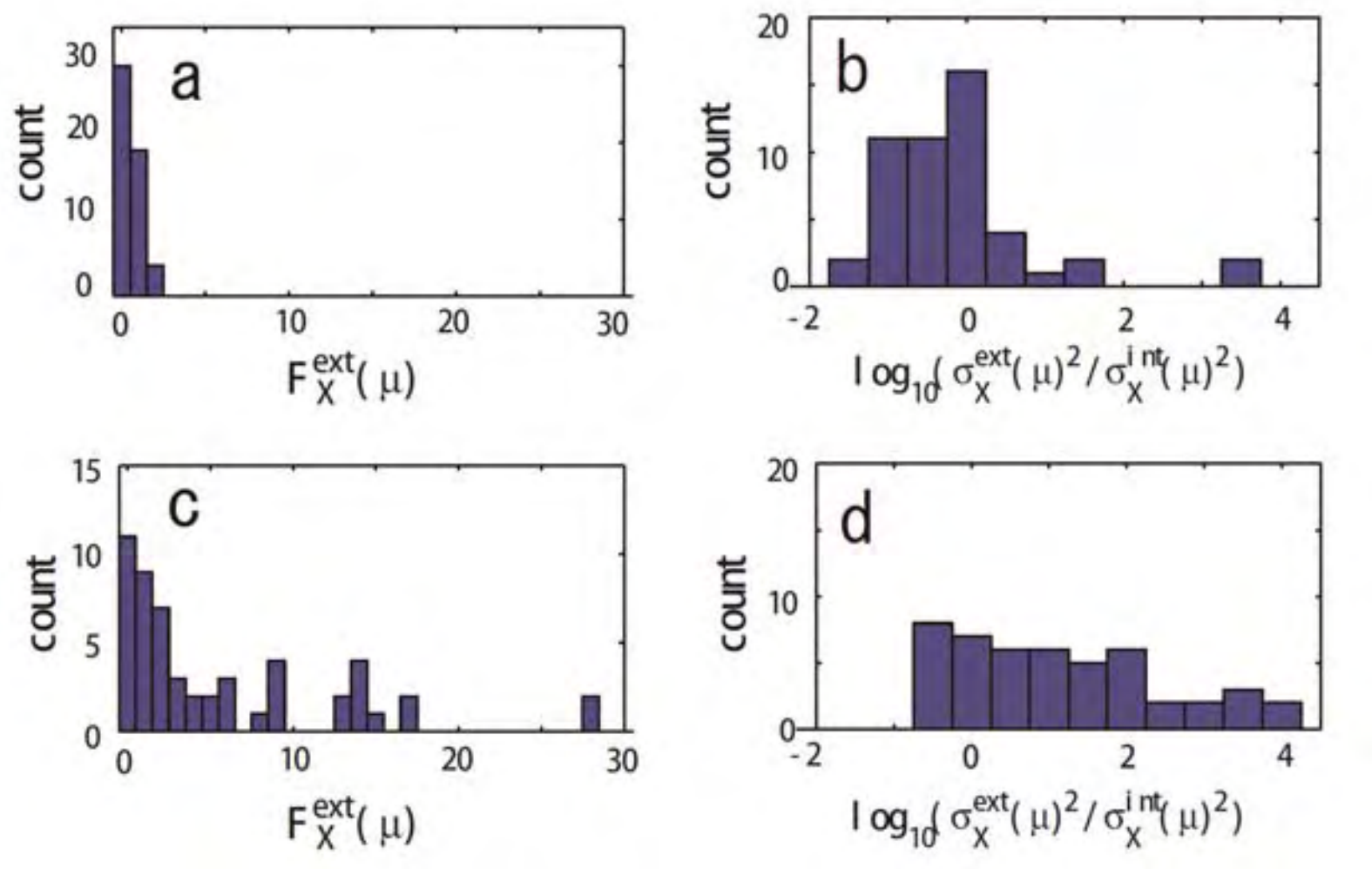

**Figure 8.** Comparison of noise between synchronous cells and asynchronous cells. **a** and **b** respectively show the distribution of $F_X^{ext}(\mu, t)$ and that of $\sigma_X^{ext}(\mu)^2/\sigma_X^{int}(\mu)^2$ of synchronous cells calculated by sampling 5000 trajectories at the same cell-cycle phases. Distributions over 125 time-points are shown. **c** and **d** show those of asynchronous cells calculated by sampling 5000 trajectories at random phases. Distributions over 100 sets of 5000 trajectories are shown. In **b** and **d**, tails of $\sigma_X^{ext}(\mu)^2/\sigma_X^{int}(\mu)^2 > 10^4$ are not shown. Distributions at $\sigma_X^{ext}(\mu)^2/\sigma_X^{int}(\mu)^2 > 10^4$ arise from proteins which have very small numbers for most of the cell-cycle duration.

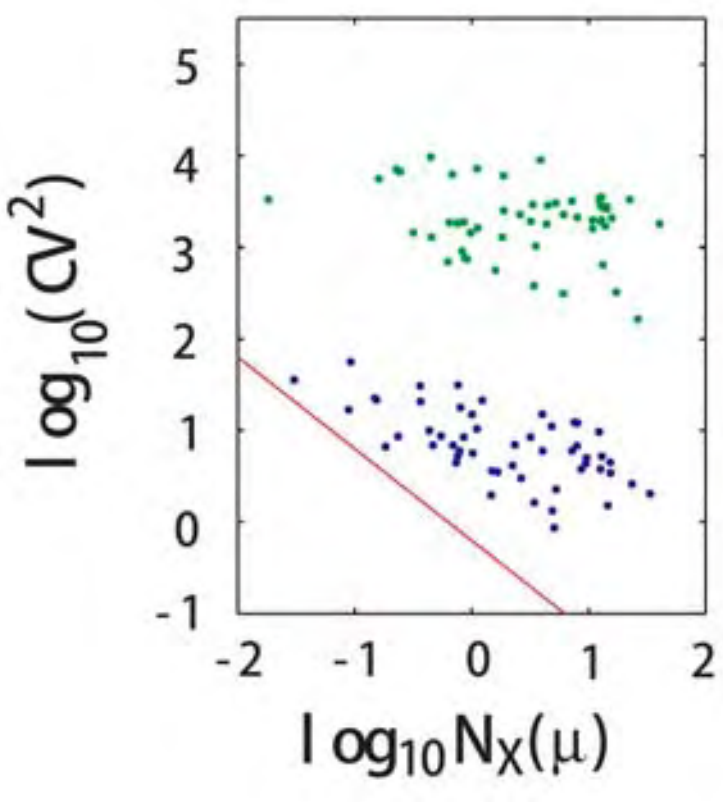

**Figure 9.** Dominance of intrinsic or extrinsic noise. CV$(\mu, X)^2$ of the number of proteins of ungated cells (green) and that of gated cells (blue) are plotted as functions of $N_X(\mu)$. Intrinsic noise is dominant in gated cells to make CV$(\mu, X)^2$ roughly proportional to $1/N_X(\mu)$. Red line has a slope of -1. The number of sampled trajectories is 100 for gated cells and 3000 for ungated cells.

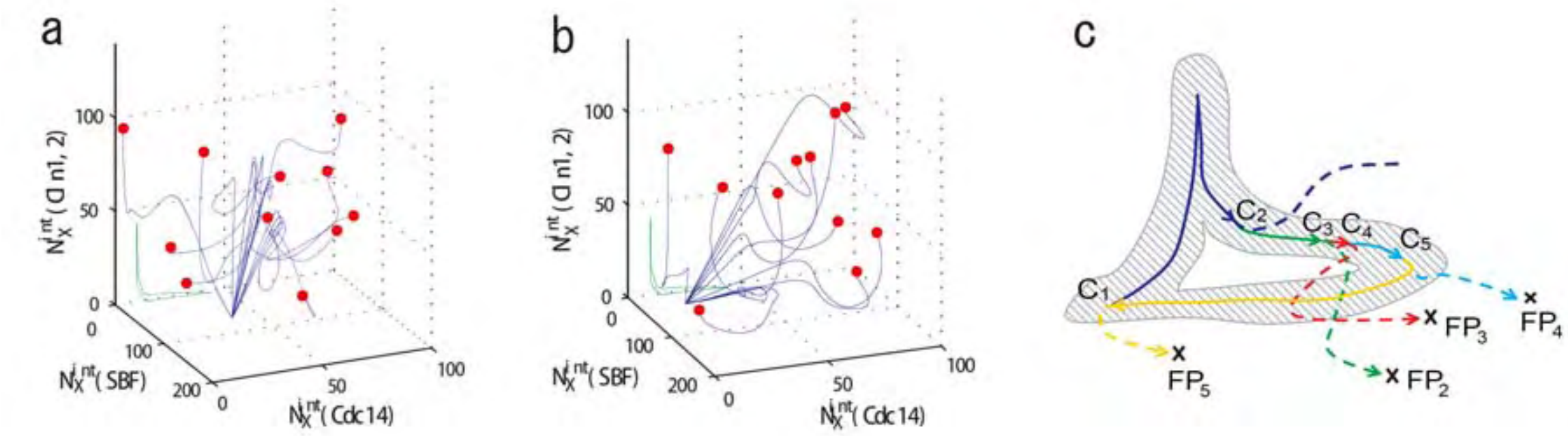

**Figure 10.** Convergence of trajectories to a fixed point. (**a**) Eleven trajectories stating from $C_3$ with random initial conditions (filled red circles) converge to $FP_3$ when stage3 is prolonged. (**b**) Eleven trajectories stating from $C_5$ with random initial conditions (filled red circles) converge to $FP_5$ when stage5 is prolonged. Blue lines are trajectories projected onto the three dimensional space of $N_X^{\mathrm{int}}$(SBF, $t$) with $X$=(0p)(1p), $N_X^{\mathrm{int}}$(Cdc14, $t$) with $X$=(outside), and $N_X^{\mathrm{int}}$(Cln1,2, $t$) with $X$=(1p)(1u). See Appendix for the definition of $X$. The green line represents the standard limit cycle. (**c**) An illustrative explanation of how the consecutively appearing fixed points drive the cell-cycle oscillation. The standard limit cycle is shown in the same three dimensional space as in **a** and **b**. Each stage in the limit cycle is specified by different colors: stage1 (dark blue), stage2 (green), stage3 (red), stage4 (light blue), and stage5 (orange). When stage$i$ is prolonged for $i$=2-5, the trajectory approaches the fixed point, $FP_i$, as shown by dashed lines. When stage1 is prolonged, trajectories tend to converge along the dashed blue line but the corresponding fixed point was not numerically found in the model. Extrinsic noise induces fluctuations of trajectories in the cell-cycle attractor which is designated by the hatched region.

## Consecutive appearance of fixed points

Though both intrinsic and extrinsic noises are large, cell cycle remains stable owing to the large basin of attraction of the cell-cycle attractor. Mechanism of attraction of trajectories to the cell-cycle attractor can be analyzed by calculating the long-time asymptotic behavior of trajectories. This behavior is examined by prolonging each stage one by one: We assume the situation that the checkpoint is so stringent or the release of Cdc14 or the cell-division is prohibited to prevent the system from passing over $C_{i+1}$. Then, the cell cycle is arrested at stage$i$. For $i$=2-5, trajectories thus arrested at stage$i$ converged to a fixed point characteristic to each stage. This fixed point corresponds to a set of constants, { $N_{m\alpha}^{\text{int}}(\mu,i)$ , $N_X^{\text{int}}(\mu,i)$ , $\sigma_{m\alpha}^{\text{int}}(\mu,i)^2$ , $\sigma_X^{\text{int}}(\mu,i)^2$ , and $D^{\text{int}}(\mu,i)$ }, and we call this set FP$_i$. In Fig.10 examples of trajectories converged to FP$_3$ and FP$_5$ are shown. Trajectories converge to FP$_i$ as $\lim_{t\to\infty}\sigma_{m\alpha}^{\text{ext}}(\mu,t)^2 = \lim_{t\to\infty}\sigma_X^{\text{ext}}(\mu,t)^2 = 0$ . $\sigma_X^{\text{ext}}(\mu,t)^2$ quickly approaches 0 when $\mu$ is the protein rapidly degraded through ubiquitination, while $\sigma_X^{\text{ext}}(\mu,t)^2$ for other proteins decreases rather slowly by taking longer time than $T_0(i)$.

The large basin of attraction of FP$_i$ is the origin of the large basin of attraction of the cell-cycle attractor. Trajectories starting from distributed initial states tend to converge toward FP$_i$. In the usual physiological condition, however, the next cellular-event of C$_{i+1}$ takes place before trajectories reach FP$_i$ and brings the system into stage$i$+1 to direct trajectories to FP$_{i+1}$. In this way, the cell-cycle oscillation is maintained by the consecutive disappearance and appearance of {FP$_i$}. It should be noted that FP$_i$ is apart from the standard limit cycle as shown in Fig.10. This deviation of fixed points allows smooth oscillations in protein and mRNA levels without being trapped at each FP$_i$. In spite of such deviation of fixed points from the standard oscillatory trajectories, shift of the fixed point from FP$_i$ to FP$_{i+1}$ is the driving force to move the system from stage$i$ to stage$i$+1. This mechanism of cell-cycle dynamics is illustrated in Fig.10c. As shown in Figs.5 and 10, width of the basin of attraction of thus generated cell-cycle attractor is $\delta N_X(\mu) > 10^2$, while as shown in Fig.9, the width $\sigma_X^{\text{total}}(\mu)$ of the region around which trajectories stochastically wander during cell cycle is $\sigma_X^{\text{total}}(\mu) \approx \sqrt{(\sigma_X^{\text{ext}}(\mu))^2 + (\sigma_X^{\text{int}}(\mu))^2} \approx 10^0\text{-}10^2$. Such large basin of attraction with $\delta N_X(\mu) > \sigma_X^{\text{total}}(\mu)$ ensures the stable oscillation in cell cycle.

## SUMMARY AND DISCUSSIONS

In this paper a stochastic model of cell cycle of budding yeast was constructed and statistical features of noise in the cell-cycle oscillation were analyzed. The model predicted that the amplitude of protein-level fluctuation is as large as $\sqrt{(\sigma^{\text{ext}})^2 + (\sigma^{\text{int}})^2 / N} \approx 10^1 - 10^0$ when an ensemble of synchronous cells are sampled. In spite of such intense stochasticity, the simulated cell cycle shows stable oscillation and attracts trajectories from widely scattered initial conditions. This stability of cell cycle is assured by consecutively appearing fixed points, each of which has a large basin of attraction. Tyson and colleagues (7, 8) showed with the deterministic model of cell cycle that the oscillation is maintained by cyclic transitions between two fixed points. In their model, transition is strongly affected by a continuous growth of the cell volume which regulates rates in the reaction network. In the present model, the reaction network is controlled by many other molecular mechanisms including check points, DNA replication, and cytokinesis, which then yield a larger number of consecutively appearing fixed points. In this sense the present model is an extension of the model of Tyson et al. toward the direction to treat the richer biochemical mechanisms to regulate the core reaction network.

The fixed-point states in the model are deviating from the usual physiological states of oscillation but appear when the lifting of checkpoints is postponed. The model showed that the hallmark of appearance of fixed points is diminution of the extrinsic noise. Comparison between the statistical features of noises at a fixed point in the model and those in the cells arrested in the corresponding stage in experiments should be important to confirm the mechanism proposed in this paper.

An interesting question is on how the perturbed cells are attracted to the cell-cycle attractor. It is left for further study to compare the simulated pathways of attraction of cells with experiments. It would be also interesting to examine whether the consecutive appearance of fixed points is the effective design principle in other reaction networks in cell as well (29, 30). Quantitative comparison of features of noisy dynamics should provide a key to examine whether such design principle works in those reaction networks.

## APPENDIX

### Chemical states of proteins

Activity and stability of individual proteins are dependent on their chemical states. For example, some proteins need to be phosphorylated to show the catalytic activity, and proteins are rapidly degraded if ubiquitinated. When a protein can be phosphorylated by a kinase, we write the chemical state of the protein as $X$ = (αp), where $\alpha = 1$ or 0, and p indicates that the chemical modification takes place on the phosphorylation site. If the phosphorylated form of the protein is active and the dephosphorylated form is inactive, we write the former as $X$ = (1p) and the latter as (0p), and if the phosphorylated form is inactive and the dephosphorylated form is active, the former is $X$ = (0p) and the latter is (1p). When a protein is targeted not only by a kinase but also by a ubiquitin ligase, then the phosphorylation site is denoted by p and the ubiquitination site is denoted by u. The chemical state is represented by $X$ = (αp)(α’u). We write α’ = 1 when the protein is not ubiquitinated, and α’ = 0 when ubiquitinated. Fig.11 describes examples of reaction schemes. Chemical states of Cdc14 are distinguished by its location whether Cdc14 is confined in the nucleolus with $X$ = (inside) or diffuses over cytoplasm with $X$ = (outside). See Table 1 for the catalog of the chemical states considered in the model.

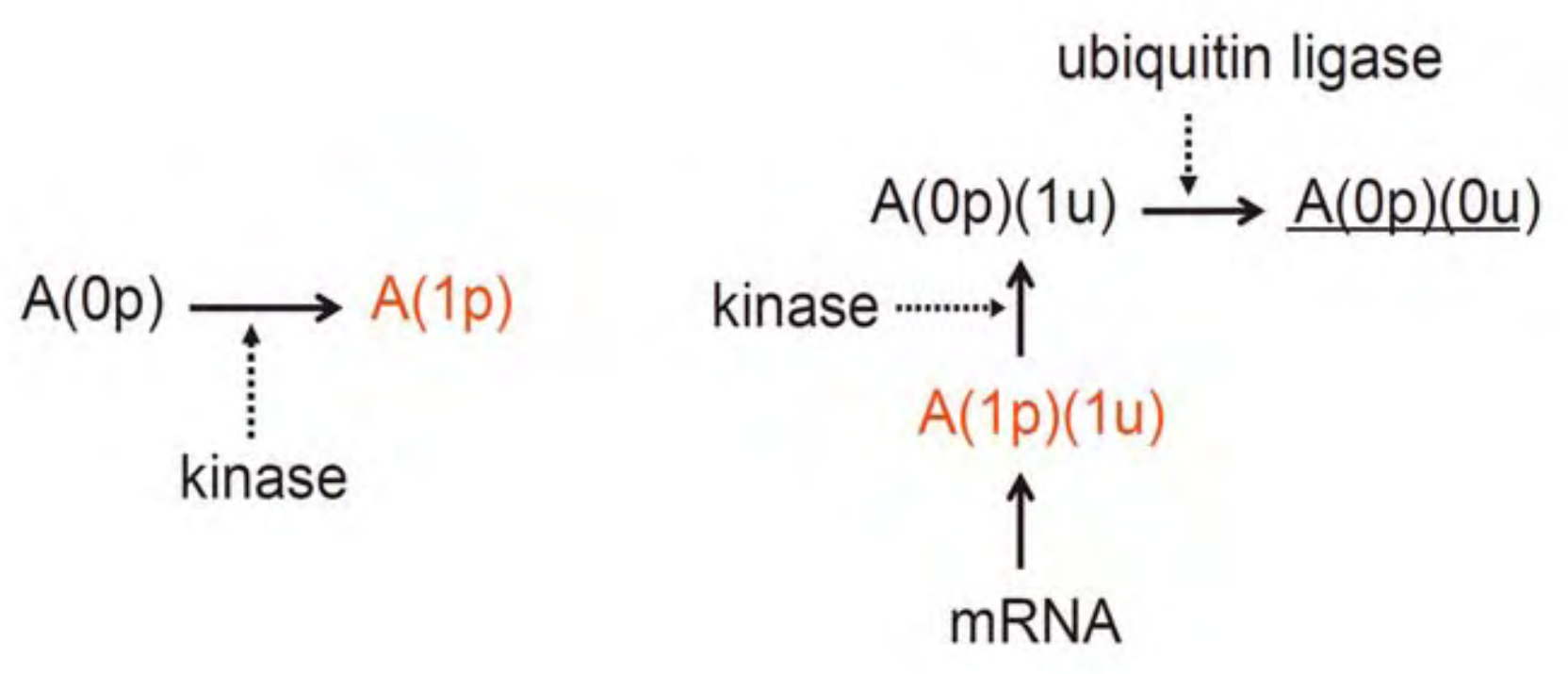

**Figure 11.** Examples of reaction schemes. (a) Phosphorylation and (b) translation, phosphorylation and ubiquitination. Protein A in the chemical state $X$ is denoted by A$X$. Catalytic actions are denoted by dotted arrows. Active proteins are denoted in red. The unstable short-lived protein is underlined. Although mRNA and all forms of proteins are assumed to be degraded with certain specific rates in the model, those degradation processes are omitted from this figure.

Table1. Activity and stability of proteins in the model

| protein | state | activitiy | stability | location |
|---|---|---|---|---|
| Cln3 | (1p)(1u) | (+) | + | nuclear |
| | (0p)(1u) | (+) | + | nuclear |
| | (0p)(0u) | (+) | – | nuclear |
| SBF | (1p)(1p) | + | + | nuclear |
| | (0p)(1p) | – | + | nuclear |
| | (0p)(0p) | – | + | cytoplasm |
| MBF | (1p)(1p) | + | + | nuclear |
| | (0p)(1p) | – | + | nuclear |
| | (0p)(0p) | – | + | cytoplasm |
| Cln1,2 | (1p)(1u) | + | + | nuclear |
| | (0p)(1u) | + | + | cytoplasm |
| | (0p)(0u) | + | – | cytoplasm |
| Sic1 | (1p)(1u) | + | + | whole |
| | (0p)(1u) | + | + | whole |
| | (0p)(0u) | + | – | whole |
| Clb5,6 | (1u) | + | + | nuclear |
| | (0u) | + | – | nuclear |
| Sic1/Clb5,6 | (1p)(1u)(1u) | – | | nuclear |
| | (0p)(1u)(1u) | – | | nuclear |
| | (0p)(0u)(1u) | – | | nuclear |
| | (1p)(1u)(0u) | – | | nuclear |
| | (0p)(1u)(0u) | – | | nuclear |
| | (0p)(0u)(0u) | – | | nuclear |
| Ndd1 | (1p)(1u) | + | + | nuclear |
| | (0p)(1u) | – | + | nuclear |
| | (1p)(0u) | + | – | nuclear |
| | (0p)(0u) | – | – | nuclear |
| Clb1,2 | (1u) | + | + | nuclear |
| | (0u) | + | – | nuclear |
| Sic1/Clb1,2 | (1p)(1u)(1u) | – | | nuclear |
| | (0p)(1u)(1u) | – | | nuclear |
| | (0p)(0u)(1u) | – | | nuclear |
| | (1p)(1u)(0u) | – | | nuclear |
| | (0p)(1u)(0u) | – | | nuclear |
| | (0p)(0u)(0u) | – | | nuclear |
| Cdc20 | (1p)(1u) | (+) | + | nuclear |
| | (0p)(1u) | (+) | + | nuclear |
| | (1p)(0u) | (+) | – | nuclear |
| | (0p)(0u) | – | – | nuclear |
| Pds1 | (1u) | + | + | nuclear |
| | (0u) | + | – | nuclear |
| Cdc14 | outside | + | outside of nucleolus | |
| | inside | – | inside of nucleolus | |
| Cdh1 | (1p)(1p)(1p) | + | + | nuclear |
| | (1p)(1p)(0p) | – | + | cytoplasm |
| | (1p)(1p)(0p) | – | + | cytoplasm |
| | (1p)(0p)(0p) | – | + | cytoplasm |
| | (0p)(1p)(1p) | – | + | cytoplasm |
| | (0p)(1p)(0p) | – | + | cytoplasm |
| | (0p)(0p)(1p) | – | + | cytoplasm |
| | (0p)(0p)(0p) | – | + | cytoplasm |
| Swi5 | (1p) | + | – | nuclear |
| | (0p) | – | + | cytoplasm |

Activity: +active, –inactive, (+) active only in specific stages.

Stability: +stable, – highly unstable.

Apart from Cdc14, "Location" is used in the model only to determine the distribution ratio in the cell separation.

## ACKNOWLEDGEMENTS

This work was supported by grants from the Ministry of Education, Culture, Sports, Science, and Technology, Japan and by grants for the 21st century COE program for Frontiers of Computational Science. Y.O. thanks the Research Fellowships for Young Scientists from the Japan Society for the Promotion of Science.

**REFERENCES**

1. Newman, J.R.S., S. Ghaemmaghami, J. Ihmels, D.K. Breslow, M. Noble, J.L. DeRisi, and J.S. Weissman. 2006. Single-cell proteomic analysis of *S. cerevisiae* reveals the architecture of biological noise. *Nature* 441:840-846.
2. Raser, J.M., and E.K. O'Shea. 2005. Noise in gene expression: origins, consequences, and control. *Science* 309:2010-2013.
3. Sigal, A., R. Milo, A. Cohen, N. Geva-Zatorsky, Y. Klein, Y. Liron, N. Rosenfeld, T. Danon, N. Perzov, and U. Alon. 2006. Variability and memory of protein levels in human cells. *Nature* 444:643-646.
4. Field, C., R. Li, and K. Oegema. 1999. Cytokinesis in eukaryotes: a mechanistic comparison. *Curr. Opin. Cell Biol.* 11:68-80.
5. Mendenhall, M.D., and A.E. Hodge. 1998. Regulation of Cdc28 cyclin-dependent protein kinase activity during the cell cycle of the yeast Saccharomyces cerevisiae. *Microbiol. Mol. Biol. Rev.* 62:1191-1243.
6. Novak, B., A. Csikasz-Nagy, B. Gyorffy, K. Nasmyth, and J.J. Tyson. 1998. Model scenarios for evolution of the eukaryotic cell cycle. *Philos. Trans. R. Soc. Lond. B. Biol. Sci.* 353:2063-2076.
7. Chen, K.C., L. Calzone, A. Csikasz-Nagy, F.R. Cross, B. Novak, and J.J. Tyson. 2004. Integrative analysis of cell cycle control in budding yeast. *Mol. Biol. Cell* 15:3841-3862.
8. Chen, K.C., A. Csikasz-Nagy, B. Gyorffy, J. Val, B. Novak, and J.J. Tyson. 2000. Kinetic analysis of a molecular model of the budding yeast cell cycle. *Mol. Biol. Cell* 11:369-391.
9. Novak, B., and J.J. Tyson. 2003. Modelling the controls of the eukaryotic cell cycle. *Biochem. Soc. Trans.* 31:1526-1529.
10. Tyson, J.J., K. Chen, and B. Novak. 2001. Network dynamics and cell physiology. *Nat. Rev. Mol. Cell Biol.* 2:908-916.
11. Borisuk, M.T., and J.J. Tyson. 1998. Bifurcation analysis of a model of mitotic control in frog eggs. *J. Theor. Biol.* 195:69-85.
12. Nasmyth, K. 1996. At the heart of the budding yeast cell cycle. *Trends Genet.* 12:405-412.
13. Li, F., T. Long, Y. Lu, Q. Ouyang, and C. Tang. 2004. The yeast cell-cycle network is robustly designed. *Proc. Natl. Acad. Sci. U S A* 101:4781-4786.
14. Cross, F.R., V. Archambault, M. Miller, and M. Klovstad. 2002. Testing a mathematical model of the yeast cell cycle. *Mol. Biol. Cell* 13:52-70.
15. Doncic, A., E. Ben-Jacob, and N. Barkai. 2006. Noise resistance in the spindle

assembly checkpoint. *Mol. Syst. Biol.* 2:2006 0027.

16. Zhang, Y., M. Qian, Q. Ouyang, M. Deng, F. Li, and C. Tang. 2006. Stochastic model of yeast cell-cycle network. *Physica D* 219:35-39.
17. Wang, J., B. Huang, X. Xia, andZ. Sun. 2006. Funneled Landscape Leads to Robustness of Cell Networks: Yeast Cell Cycle. *PLoS Comput. Biol.* 2:1385-1394.
18. Blake, W.J., M. Kærn, C.R. Cantor, and J.J. Collins. 2003. Noise in eukaryotic gene expression. *Nature* 422:633-637.
19. Raser, J.M., and E.K. O'Shea. 2004. Control of stochasticity in eukaryotic gene expression. *Science* 304:1811-1814.
20. Elowitz, M.B., A.J. Levine, E.D. Siggia, and P.S. Swain. 2002. Stochastic gene expression in a single cell. *Science* 297:1183-1186.
21. Guido, N.J., X. Wang, D. Adalsteinsson, D. McMillen, J. Hasty, C.R. Cantor, T.C. Elston, and J.J. Collins. 2006. A bottom-up approach to gene regulation. *Nature* 439:856-860.
22. Spellman, P.T., G. Sherlock, M.Q. Zhang, V.R. Iyer, K. Anders, M.B. Eisen, P.O. Brown, D. Botstein, and B. Futcher. 1998. Comprehensive identification of cell cycle-regulated genes of the yeast *Saccharomyces cerevisiae* by microarray hybridization. *Mol. Biol. Cell* 9:3273-3297.
23. Melo, J., and D. Toczyski. 2002. A unified view of the DNA-damage checkpoint. *Curr. Opin. Cell Biol.* 14:237-245.
24. Bi, E., P. Maddox, D.J. Lew, E.D. Salmon, J.N. McMillan, E. Yeh, and J.R. Pringle. 1998. Involvement of an actomyosin contractile ring in *Saccharomyces cerevisiae* cytokinesis. *J. Cell Biol.* 142:1301-1312.
25. Shaw, S.L., P. Maddox, R.V. Skibbens, E. Yeh, E.D. Salmon, and K. Bloom. 1998. Nuclear and spindle dynamics in budding yeast. *Mol. Biol. Cell* 9:1627-1631.
26. Yeh, E., R.V. Skibbens, J.W. Cheng, E.D. Salmon, and K. Bloom. 1995. Spindle dynamics and cell cycle regulation of dynein in the budding yeast, *Saccharomyces cerevisiae*. *J. Cell Biol.* 130:687-700.
27. Strunnikov, A.V. 2003. Condensin and biological role of chromosome condensation. *Prog. Cell Cycle Res.* 5:361-367.
28. Griffin, T.J., S.P. Gygi, T. Ideker, B. Rist, J. Eng, L. Hood, and R. Aebersold. 2002. Complementary profiling of gene expression at the transcriptome and proteome levels in *Saccharomyces cerevisiae*. *Mol. Cell Proteomics* 1:323-333.
29. Sasai, M., and P.G. Wolynes. 2003. Stochastic gene expression as a many-body problem. *Proc. Natl. Acad. Sci. U S A* 100:2374-2379.
30. Wang, J., B. Huang, X. Xia, and Z. Sun. 2006. Funneled landscape leads to

robustness of cellular networks: MAPK signal transduction. *Biophys. J.* 91:L54-56.

31. Gillespie, D. T. 1977. Exact stochastic simulation of coupled chemical reactions. *J. Phys. Chem.* 81:2340-2361.

**Supporting Text1**
**Stable stochastic dynamics in yeast cell cycle**
**Yurie Okabe and Masaki Sasai**

In this Supporting Text, reactions involved in Fig.1 of the main text are explained. Fig.1 contains reactions among 13 genes, 13 mRNAs, and 53 chemical states of proteins and protein complexes. Although all mRNAs and all forms of proteins are assumed to be degraded with certain specific rates in the model, explanation of those degradation processes is omitted in the following description. For figures of chemical schemes inserted in this Supporting Text, protein A in the chemical state *X* is denoted by A*X*. Catalytic actions are denoted by dotted arrows. Active proteins are denoted in red. The unstable short-lived protein is underlined.

## 1) Cln3

Experimental observations

The *CLN3* promoter contains ECB (early cell cycle box) and the Swi5 binding site [1, 2]. Although the *CLN3* mRNA level increases three- to four-fold at around the M-G1 boundary [3], the Cln3 protein level is kept low and oscillation of the Cln3 level is modest throughout the cell cycle [4]. Cln3 localizes to nucleus [5] and forms Cln3/Cdc28 complex. The phosphorylated Cln3 is ubiquitinated in a Cdc34-dependent manner [6-8] and the ubiquitinated Cln3 is highly unstable with a half-life time of ~10 min [4, 8].

Model

The *CLN3* expression is assumed to be regulated by the transcriptional activator Swi5. We assume the complex, Cln3/Cdc28, autophosphorylates itself. $PX_1$ represents the ubiquitin ligase working on Cln3, whose abundance is assumed to be constant in the model. Thus, the phosphorylated Cln3 denoted by Cln3(0p)(1u) is ubiquitinated with a constant rate. All forms of Cln3 can work on SBF and MBF during stage1.

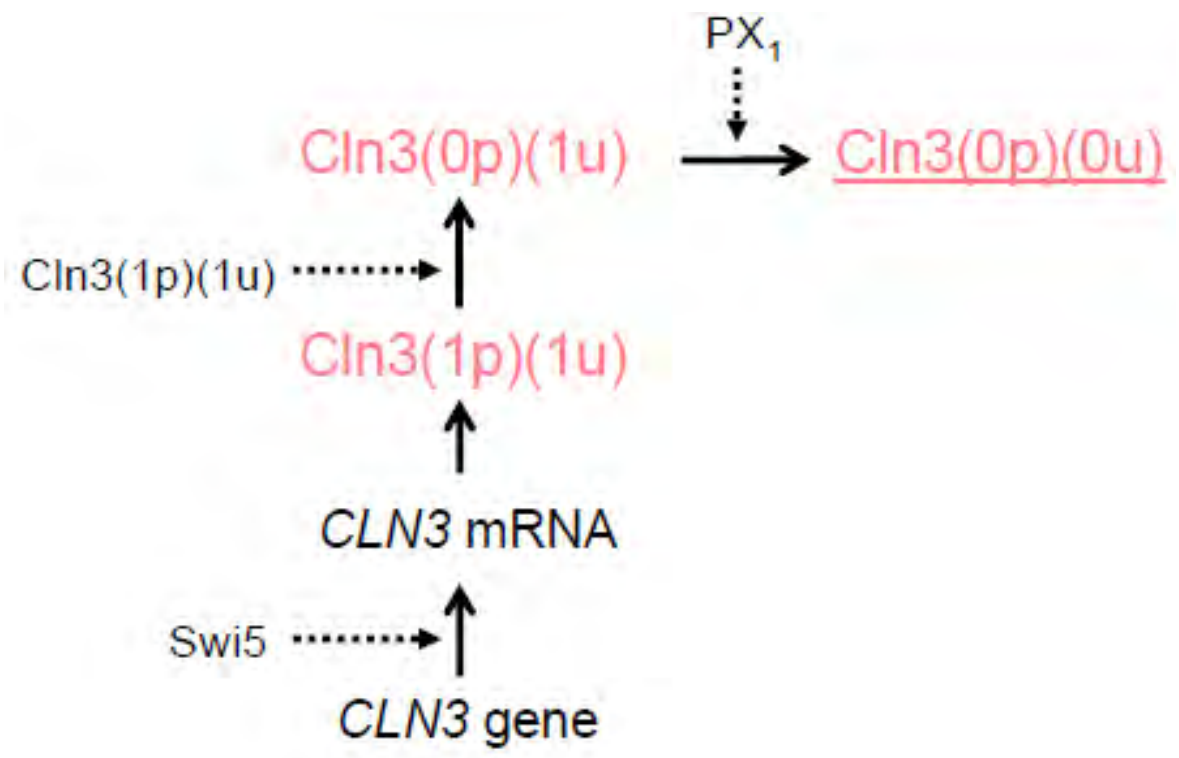

**2) SBF**

Experimental observations

SBF (SCB binding factor) is a transcriptional activator composed of Swi4 and Swi6, and binds to the SCB sequence in the form of a heterodimer [9]. Abundance of SBF changes through the cell cycle partially because of the fluctuation in the *SWI4* mRNA level, but this change is not much correlated to its ability to regulate the *CLN2* transcription [10]. Prior to late G1, SBF binds to the SCB promoter, but Whi5 binds to SBF at the promoter and inhibits the SBF activity. In late G1, Cln3/Cdc28 promotes dissociation of Whi5 from SBF at the promoter and thereby SBF recovers its activity [11]. In G2-M phase, SBF dissociates from the promoter when Swi4 is phosphorylated by Clb1,2/Cdc28 [10, 12]. The nuclear localization of Swi6 is regulated in a cell cycle dependent manner [13], whereas the DNA binding component, Swi4, remains in nucleolus throughout the cell cycle [14]. Phosphorylation of Swi6 by cyclin/Cdc28 at the end of G1 prevents nuclear localization of Swi6. In late M, Cdc14 is released from nucleolus and dephosphorylates Swi6, which leads to the accumulation of Swi6 in nucleus [13].

Model

We treat the SBF complex as a single unit and do not take account of its individual components separately. SBF is assumed to be constantly produced from the putative SBF gene and its mRNA. SBF has two symbolic phosphorylation sites denoted by ($\alpha$p) and ($\alpha$'p'). While ($\alpha$p) represents change of the chemical state in the G1/S transition, ($\alpha$'p') represents that in G2 and M phases. $\alpha$ is turned to be 1 by the action of Cln3/Cdc28 but the period that Cln3/Cdc28 is active is limited only to stage1. $PX_2$ represents the hypothetical inactivator of SBF(1p)(1p'), whose amount is assumed to be constant throughout the cell cycle. Reactions on (α'p') represent changes in both Swi4 and Swi6. We assume phosphorylation of Swi6 is carried out mainly by Clb1,2/Cdc28 rather than by Clb5,6/Cdc28, so that the (α'p')-site is phosphorylated by Clb1,2/Cdc28 and dephosphorylated by Cdc14.

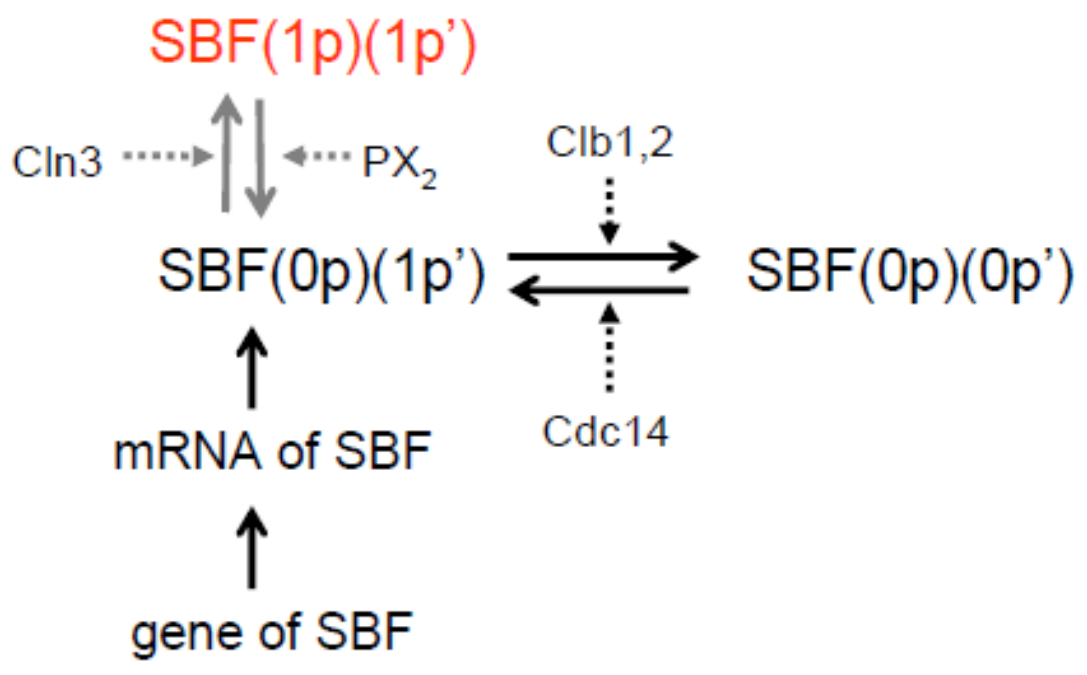

### 3) MBF

Experimental observations

MBF (MCB binding factor) is a transcriptional activator composed of Mbp1 and Swi6, which binds to the MCB sequence in the form of a single heterodimer [9]. Not much is know about the regulation of Mbp1 in the MBF complex. Swi6 is regulated as in the case of SBF complex.

Model

The model for molecular interactions of MBF is similar to that of SBF. We assume that the MBF complex is produced from the putative MBF gene and mRNA. MBF is assumed to have two reaction sites as in the case of SBF, but the (α'p')-site is phosphorylated by Clb5,6/Cdc28 instead of Clb1,2/Cdc28.

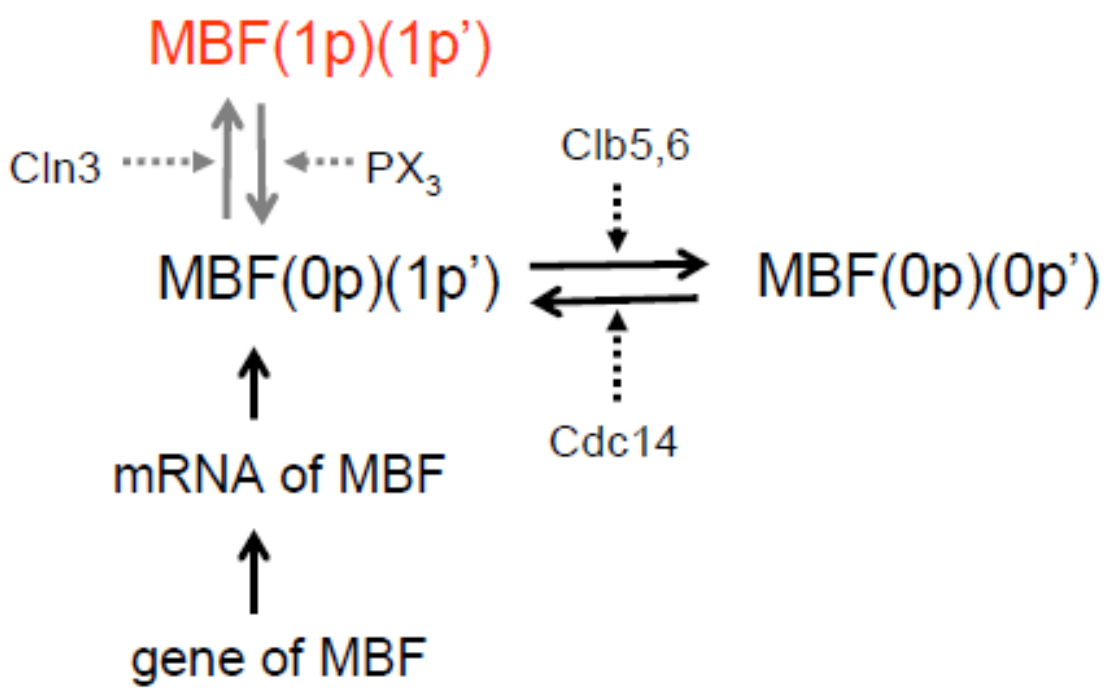

**4) Cln1,2**

Experimental observations

Expression of *CLN1* and *CLN2* is regulated by the MCB (MluI cell cycle box) and SCB (Swi4,6-depnendent cell cycle box) promoters, and the MCB and SCB promoters are activated by MBF and SBF, respectively [2, 15-17]. Clb6/Cdc28 negatively regulates the Cln2 function at the protein level [18]: Cdc28 phosphorylates both Cln1 and Cln2, and the phosphorylated Cln1 and Cln2 are ubiquitinated by $SCF^{Grr1}$ [19-21]. The ubiquitinated Cln1 and Cln2 are rapidly degraded with half-life time of 8-10 min [4, 21]. Cln2 can also form Cln2/Cdc28 complex even when Cln2 is phosphorylted by Cdc28 [21]. Although Cln2 is found at similar concentrations in cytoplasm and nucleus [22], the hypophosphorylated Cln2/Cdc28 is mainly in nucleus and the phosphorylated Cln2/Cdc28 is localized to cytoplasm [5, 23].

Model

Both SBF and MBF activate the expression of *CLN1,2*. Cln1,2 is assumed to have two reaction sites, (αp) and (αu). Clb5,6/Cdc28 phosphorylates the (αp)-site of Cln1,2, and the phosphorylated form of Cln1,2 is ubiquitinated. $PX_4$ represents the ubiquitin ligase activity of $SCF^{Grr1}$, whose abundance is assumed to be constant.

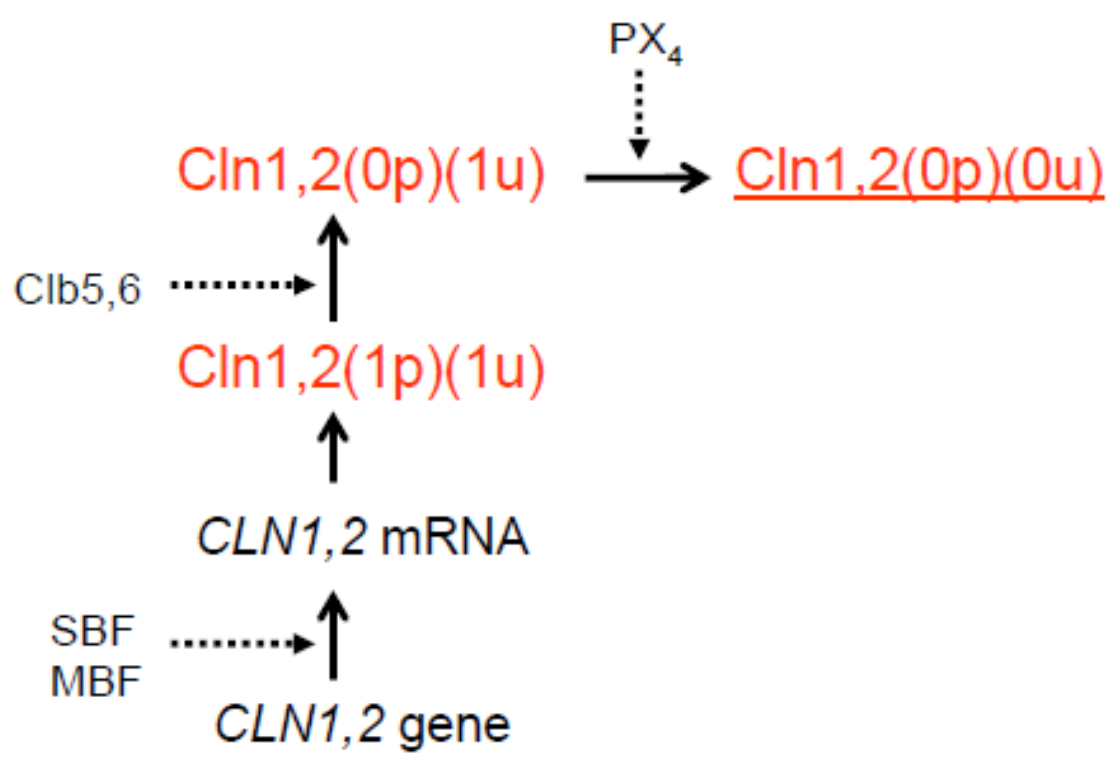

**5) Sic1**

Experimental observations

The *SIC1* promoter is activated by Swi5 and its expression increases three- to four-fold around the M-G1 boundary [24-26]. Sic1 is distributed in both cytoplasm and nucleus at similar concentrations [22], and it inhibits the Clb5/Cdc28 kinase activity during G1 by forming the ternary complex with Clb5/Cdc28 [26]. Abundance of Clb5 begins to increase at the G1-S transition, and when Clb5 exists in excess, Clb5/Cdc28 phosphorylates Sic1 [20]. Cln2/Cdc28 also phosphorylates both the monomeric Sic1 and Sic1 in the Sic1/Clb5/Cdc28 ternary complex [20, 27]. When either form of Sic1 is phosphorylated on at least six out of nine CDK sites, it is recognized and ubiquitinated by SCF$^{Cdc34}$ [20, 25-28]. Sic1 is unstable in S phase with a half-life of 10 min or less [29] and its abundance is low before the M-G1 boundary. Swi5 activates the *SIC1* expression and Cdc14 desphorylates Sic1 to avoid ubiquitination [30, 31].

<u>Model</u>

Transcription of *SIC1* is activated by Swi5. Sic1 is assumed to have two reaction sites, ($\alpha$p) and ($\alpha$u). Cln1,2, Clb1,2, and Clb5,6 phosphorylate the ($\alpha$p)-site of Sic1, which is in turn dephosphorylated by Cdc14. The phoshorylated form of Sic1, Sic1(0p)(1u), is ubiquitinated in proportion to its abundance. $PX_5$ represents the constant ubiquitin ligase activity of SCF$^{Cdc34}$. All forms of Sic1 proteins can bind to Clb1,2/Cdc28 and Clb5,6/Cdc28.

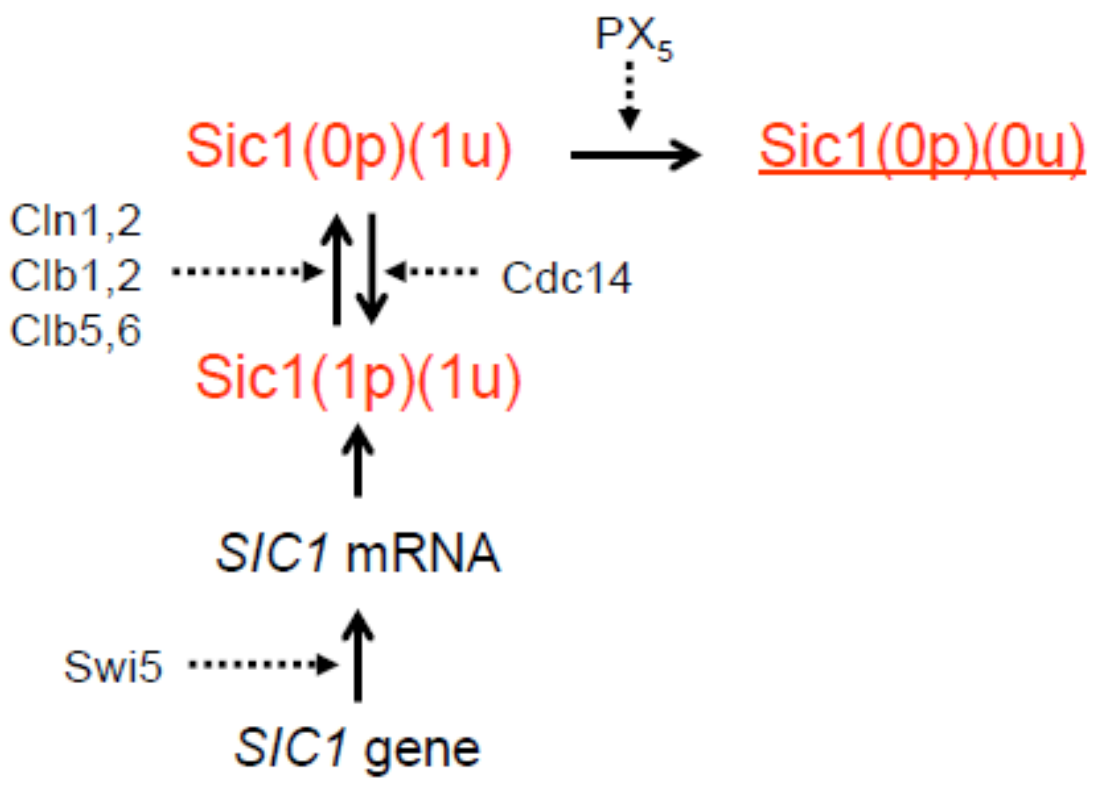

**6) Clb5,6**

<u>Experimental observations</u>

*CLB5* mRNA is very rare in early G1 and accumulates to high level, and then rapidly decreases in G2 [26]. MBF is a potential activator of *CLB5* and *CLB6*, which shows high affinity to their promoters in microarray experiments [2, 15]. During G1, abundance of Clb5/Cdc28 is low and its activity is inhibited by the association with Sic1. Clb5/Cdc28 accumulates in nucleus to increase its activity as cell enters S phase, but APC$^{Cdc20}$ leads to its sudden decrease at the metaphase-anaphase transition [26, 32]. Half-life of Clb5 is 5-10 min in G1 and 15-20 min in S and M [33].

Model

Expression of *CLB5,6* is positively regulated by MBF. Clb5,6 has a reaction site which can be ubiquitinated by Cdc20 in stage3-5. Kinase activity of Clb5,6/Cdc28 is inhibited when bound to Sic1 and the activity is recovered when Sic1 in the ternary complex is degraded.

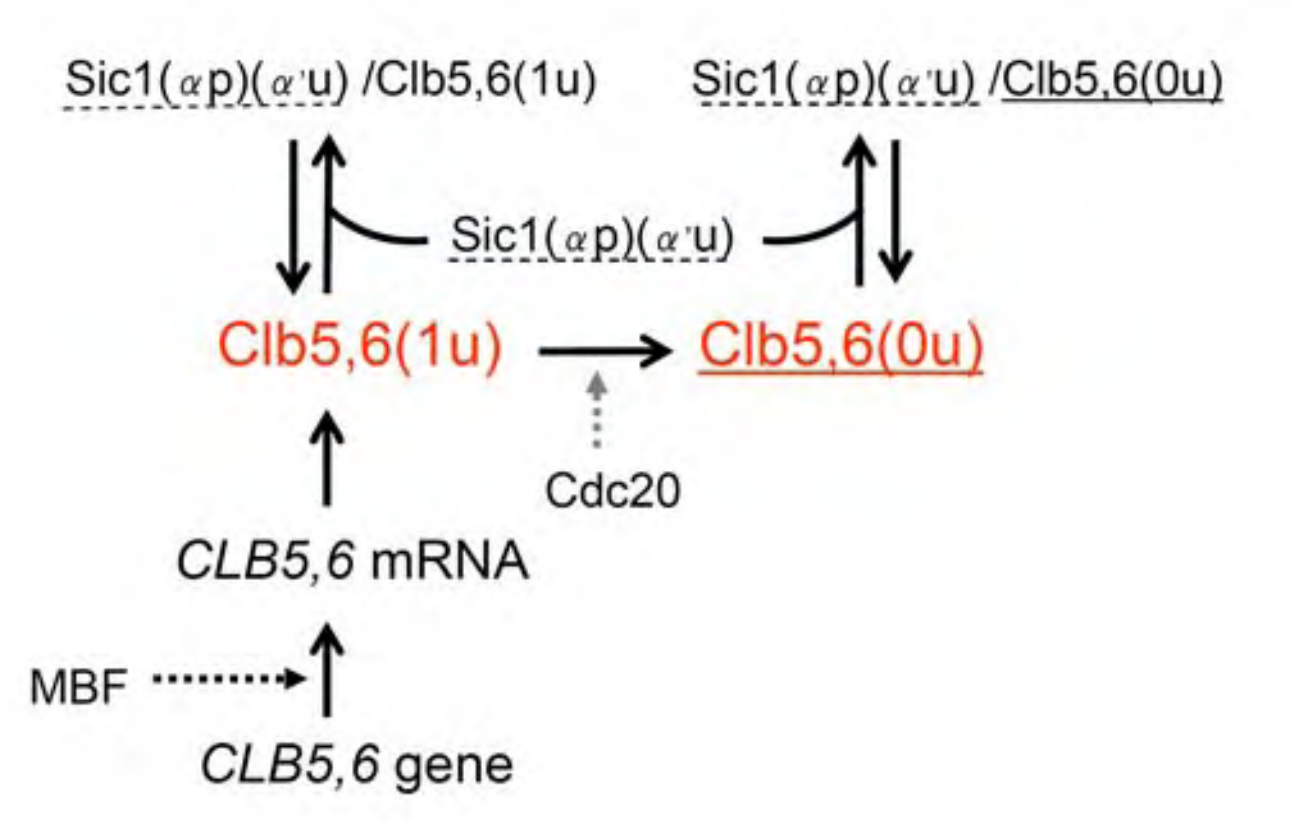

**7) Clb1,2**

Experimental observations

Transcription of *CLB2* is activated by Mcm1/Fkh2/Ndd1 during G2 and M. The microarray analyses suggest that SBF is another activator of *CLB2* [2, 16]. Clb2 is strongly localized in nucleus at all stages of cell cycle [34], but its abundance is regulated by both transcriptional activation and $APC^{Cdh1}$-mediated ubiquitination. During G1 phase, when the $APC^{Cdh1}$ level is high, Clb2 is highly unstable and barely detected. The Clb2 level begins to increase in S phase, peaks during M phase, and declines at some time in late anaphase [35-39]. Clb2 is stable during S and M with half-life of > 1h, but extremely short-lived in G1 with half-life of < 5 min [33, 40].

Model

Expression of *CLB1,2* is activated by both SBF and Ndd1. Clb1,2 is ubiquitinated by Cdh1. Clb1,2/Cdc28 is inactivated by forming a complex with Sic1 and is activated when Sic1 in the complex is degraded.

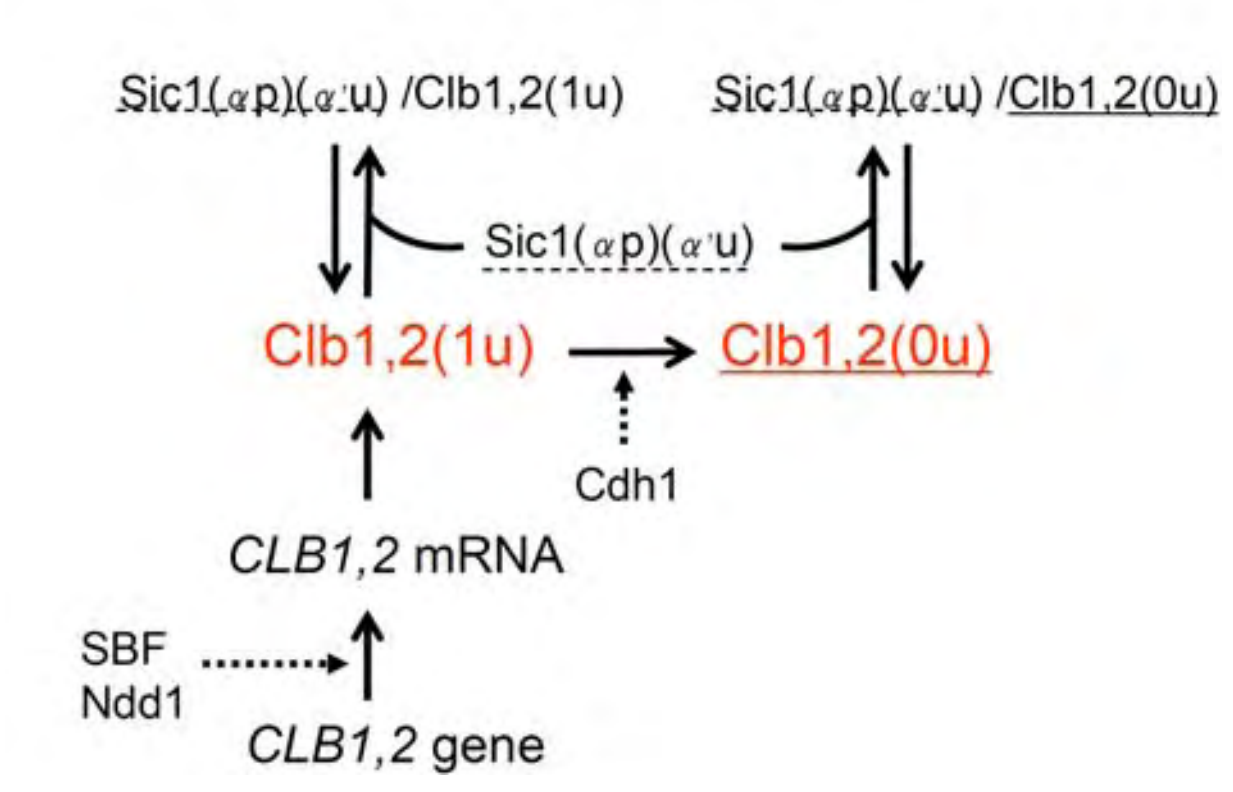

**8) Ndd1**

Experimental observations

SBF binds to and activates the *NDD1* promoter [2, 15]. Ndd1 is localized in nucleus [36] forms the Mcm1/Fkh2/Ndd1 ternary complex. The Mcm1/Fkh2 complex occupies *CLB2* and *SWI5* promoters throughout the cell cycle [41] and these promoters are activated when Ndd1 is recruited [42, 43]. For this recruitment, phosphorylation of Ndd1 by Clb2/Cdc28 is required. Ndd1 begins to decrease at the beginning of disassembly of the mitotic spindles and remains unstable until anaphase spindles disappear [36].

Model

Expression of *NDD1* is activated by SBF. We assume Ndd1 is ubiquitinated by Cdc20 and it has two reaction sites, (αp) and (αu). The former is phosphorylated by Clb1,2/Cdc28 to be active, and the latter is ubiquitinated by Cdc20 during stage3-5.

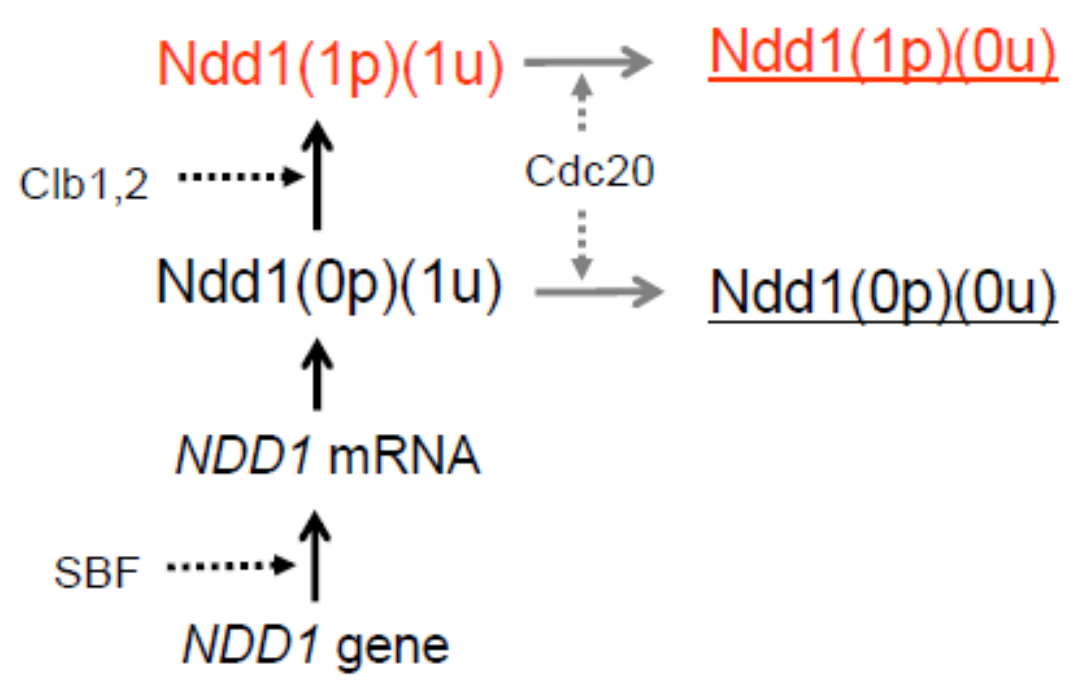

**9) Cdc20**

Experimental observations

Expression of *CDC20* is activated by the Mcm1/Fkh2/Ndd1 complex and others [2, 16], leading to the oscillation of the *CDC20* mRNA level which peaks in M phase [38]. The abundance of Cdc20 fluctuates throughout the cell cycle, rising in S phase, being maximal during M phase, and declining on exit from M phase [37, 38, 44]. Cdc20 is localized to nucleus [44] and the spindle checkpoint keeps the Cdc20 level low until all kinetochores attach to spindle microtubules. The check-point induced Cdc20 degradation requires the physical interaction between Cdc20 and Mad2 and involves APC [45]. Cdh1 is not required for this process [45], but $APC^{Cdh1}$ contributes to the degradation of Cdc20 in late G1 [46]. In this manner Cdc20 is unstable throughout the cell cycle with half-life time of $< 3$ min during S, G2, and early M and is less stable in anaphase and G1 [38]. Activity of Cdc20 is also regulated by the spindle checkpoint. Spindle checkpoint proteins, Mad2 and Mad3, bind to Cdc20 to prevent it from activating APC until the metaphase-anaphase transition [45]. Phosphorylation of the APC core subunits by Clb2/Cdc28 enhances the association of Cdc20 with APC and increases the $APC^{Cdc20}$ activity [47, 48].

Model

Expression of *CDC20* is activated by Ndd1. It is experimentally known that phosphorylation of APC by Clb2/Cdc28 promotes $APC^{Cdc20}$ activity, and this effect is represented as phosphorylation and activation of Cdc20 by Clb2/Cdc28 in the model. In order to include the effect of the checkpoint-induced Cdc20 degradation into the model, we assume that Cdc20 in the model autoubiquitinates itself throughout the cell cycle. The checkpoint represses the Cdc20 activity until the metaphase-anaphase transition takes place. We express this checkpoint mechanism by imposing the condition that Cdc20 works on the target proteins other than itself only during stage3-5: Cdc20-dependent ubiquitination of Clb5,6, Ndd1, and Pds1 is limited to stage3-5.

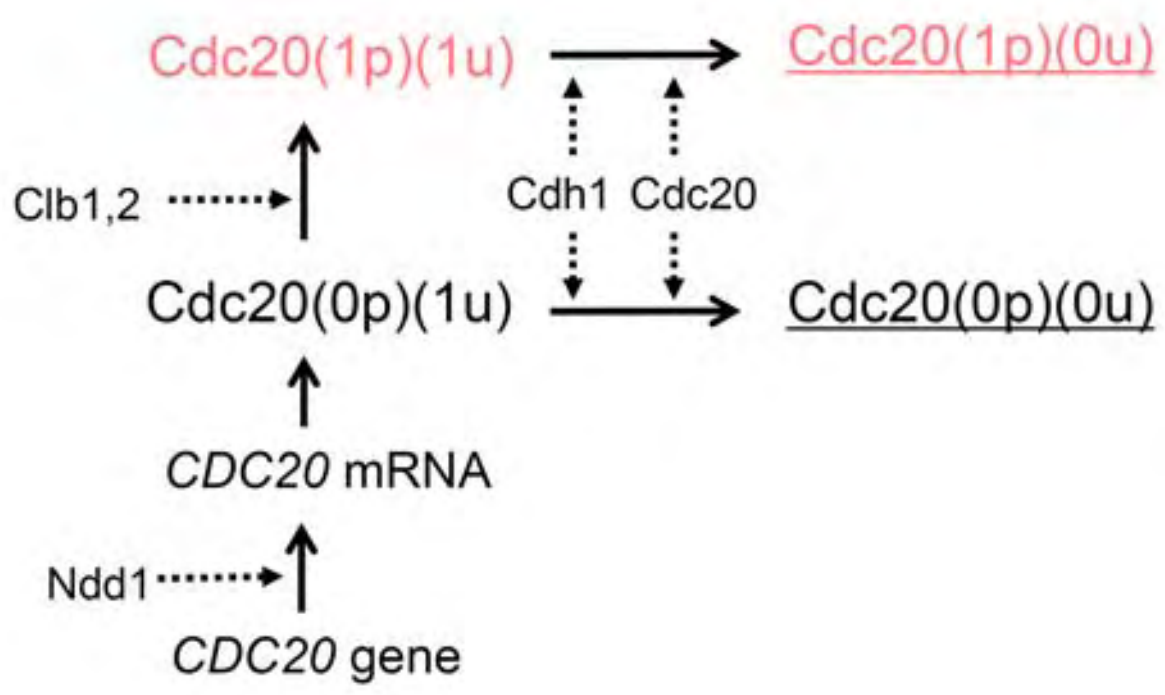

**10) Pds1**

Experimental observations

The *PDS1* mRNA level fluctuates in a cell cycle-dependent manner with maximal accumulation around the G1-S transition [49]. The microarray analysis showed that MBF binds to the *PDS1* promoter, suggesting that the expression of *PDS1* is activated by MBF [15]. Pds1 is localized in nuclear [50] and both its productivity and stability are regulated during cell cycle. At the end of metaphase, Pds1 is ubiquitinated by $APC^{Cdc20}$ and thereby undergoes rapid degradation [39, 47, 50, 51]. Pds1 is also targeted by $APC^{Cdch1}$ and is highly unstable during anaphase and G1 (half-life <15 min) [48, 51]. In consequence, Pds1 exists during the period from late G1 or from early S to the metaphase-anaphase transition [50]. The anaphase inhibitor Pds1 (securin) binds to Esp1 (separin) and inhibits the activity of Esp1. The rapid degradation of Pds1 at the metaphase-anaphase transition is required for the liberation of Eps1. When released form Pds1, Eps1 can induce cleavage of the cohesion complex that holds sister chromatids together.

Model

Expression of *PDS1* is activated by MBF. Pds1 is ubiquitinated by $APC^{Cdc20}$ (during stage3-5) and by $APC^{Cdch1}$.

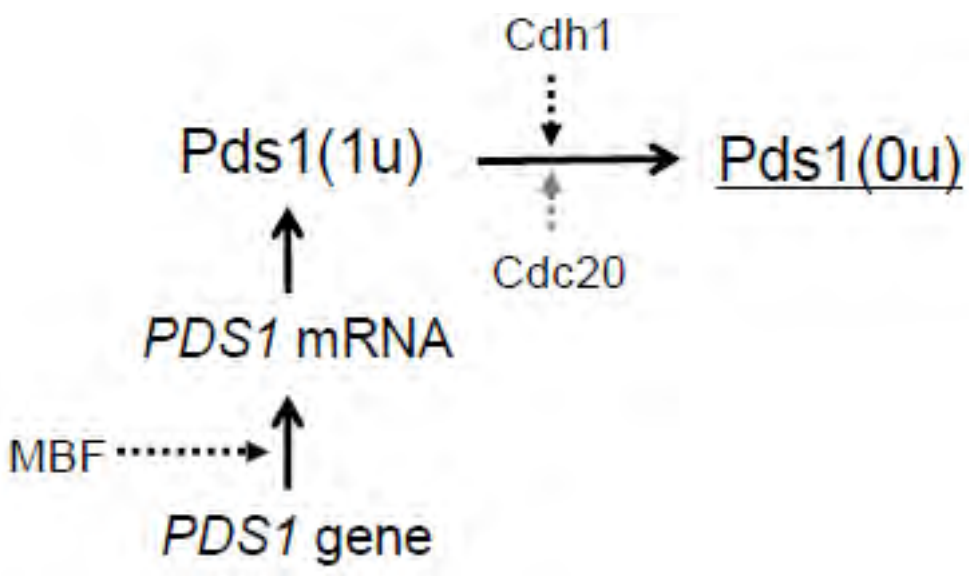

**11) Cdc14**

Experimental observations

Although the Cdc14 level is roughly constant, the subcellular localization of Cdc14 changes remarkably in a cell cycle-dependent manner. From G1 to early M phase, Cdc14 is localized in nucleolus as a part of the RENT complex, which prevents Cdc14 from phosphorylating its target proteins. Cdc14 is released from the RENT complex at some time in anaphase [35, 52]. Release of Cdc14 requires degradation of Pds1 by $APC^{Cdc20}$ [32]. The released Cdc14 spreads throughout nucleus and cytoplasm and it dephosphorylates Cdh1, Swi5, and Sic1 to promote exit from mitosis [30, 31, 53, 54]. Then, Cdc14 comes back into nucleolus as cell enters G1 phase [35]. The localization of Cdc14 is regulated by proteins which are not included in the present model.

Model

Cdc14 is distinguished by its location. Localization is regulated by changing the rates of exporting and importing Cdc14 from and to nucleolus in a stage-dependent manner: During stage4, the exporting rate of Cdc14 is $r_{ex} = (\ln 2/160)(\Delta n)n_{\mathrm{in}}$, where $n_{\mathrm{in}}$ is the number of Cdc14 locating inside of nucleolus and $\Delta n$ is the number of Pds1 molecules degraded during stage3, and the importing rate of Cdc14 is $r_{im} = 0.2(\ln 2/10)n_{\mathrm{out}}$, where $n_{\mathrm{out}}$ is the number of Cdc14 locating outside of nucleolus. During other stages, $r_{ex} = 0.2(\ln 2/160)n_{\mathrm{in}}$ and $r_{im} = (\ln 2/10)n_{\mathrm{out}}$.

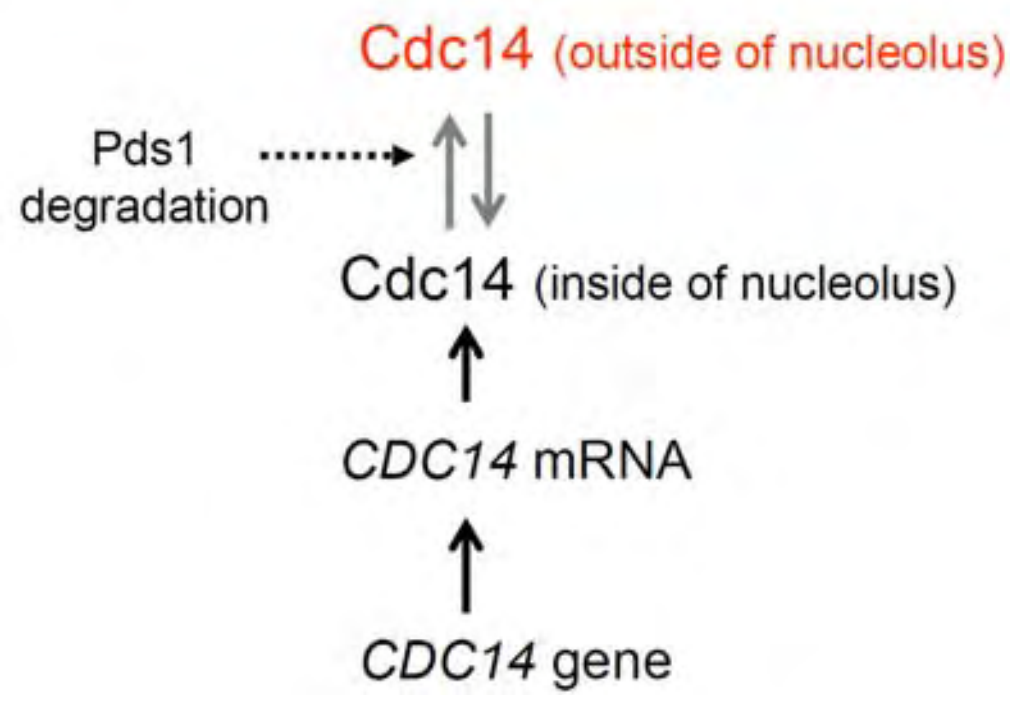

**12) Cdh1**

Experimental observations

Abundance of Cdh1, as well as that of the *CDH1* mRNA, are roughly constant during cell cycle [37, 38], but the activity of $APC^{Cdch1}$ is regulated by cyclin/Cdc28 complexes through phosphorylation of Cdc14. During S, G2, and M phases, Cdc28 associated with Cln1, Cln2, Clb1, and Clb5, phosphorylates multiple sites of Cdh1 [44, 46, 47, 55]. The phosphorylated Cdh1 do not bind to APC and is exported into cytoplasm [44]. Activity of $APC^{Cdch1}$ is restored, when Cdh1 is dephosphorylated by Cdc14 at the end of mitosis [53].

Model

We assume Cdh1 has three reaction sites, each of which is phosphorylated by a single kind of cyclin/Cdc28. Namely, Cln1,2, Clb1,2 and Clb5,6 respectively work on different sites of Cdh1. All these reaction sites are dephosphorylated by Cdc14 independently. Among eight forms of Cdh1, only Cdh1(1p)(1p)(1p) is assumed to be active.

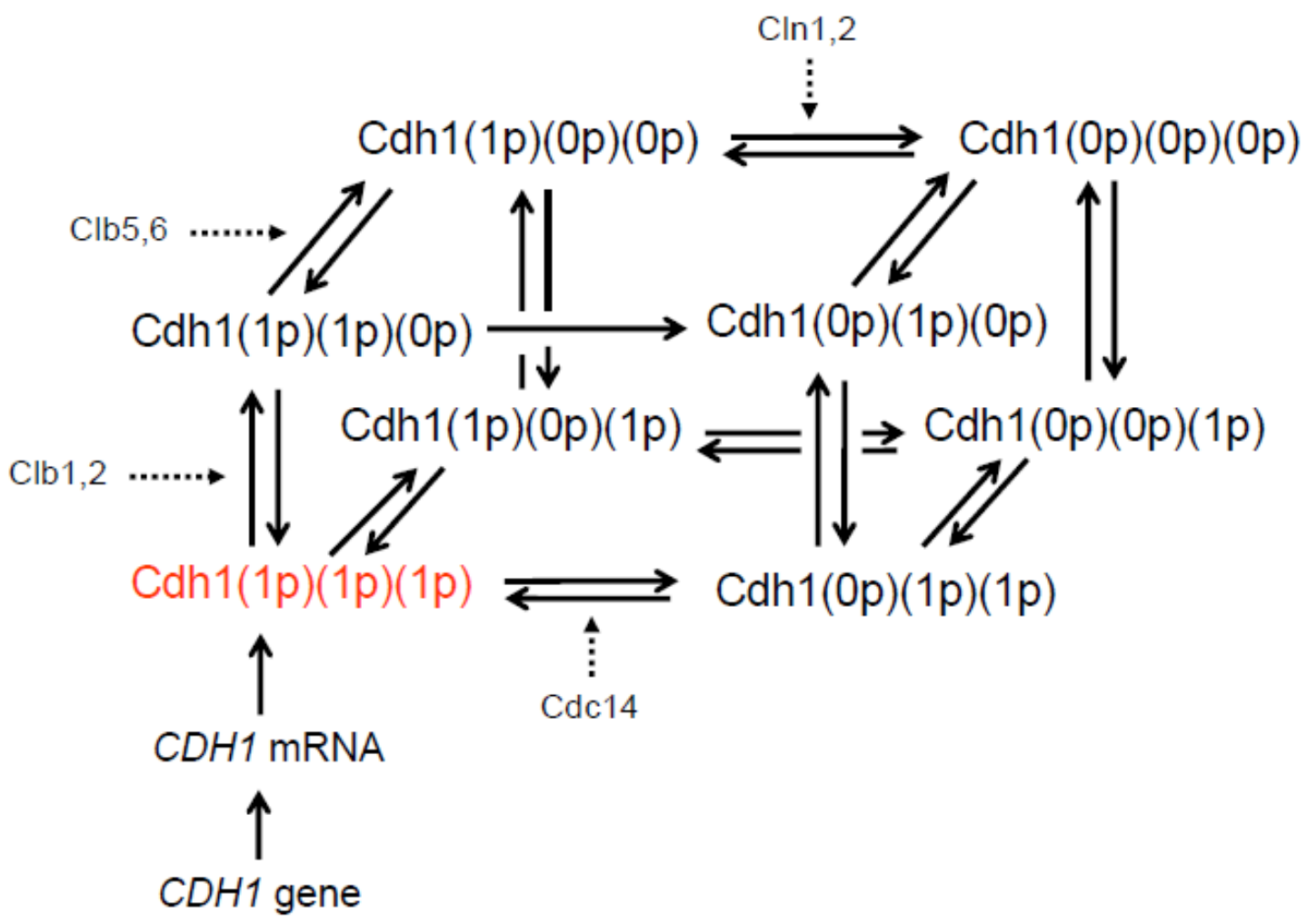

**13) Swi5**

Experimental observations

Transcription of *SWI5* is specific to G2 and M phases. Swi5 is the transcriptional factor of Sic1 and Cln3, and the activity of Swi5 is regulated by its localization. Prior to anaphase, Cdc28 mediates the Swi5 localization in cytoplasm [54, 56]. Around the anaphase-telophase boundary, Swi5 is dephosphorylated by Cdc14 and accumulates in nucleus [54, 57]. Swi5 is highly unstable in the nucleus and the majority of Swi5 is degraded by the time of cell separation [57].

Model

Expression of *SWI5* is activated by MBF. Swi5 is phosphorylated by Clb1,2/Cdc28 and dephosphorylated by Cdc14. The phosphorylated form of Swi5 is assumed to localize in cytoplasm and hence be inactive. The dephosphorylated form of Swi5, on the other hand, is assumed to localize in nucleus, where it can activate the transcription of *SIC1* and *CLN3*. We assume dephosphorylated form of Swi5 is rapidly degraded with the same half-life time of ubiquitinated proteins because Swi5 is highly unstable in nucleus.

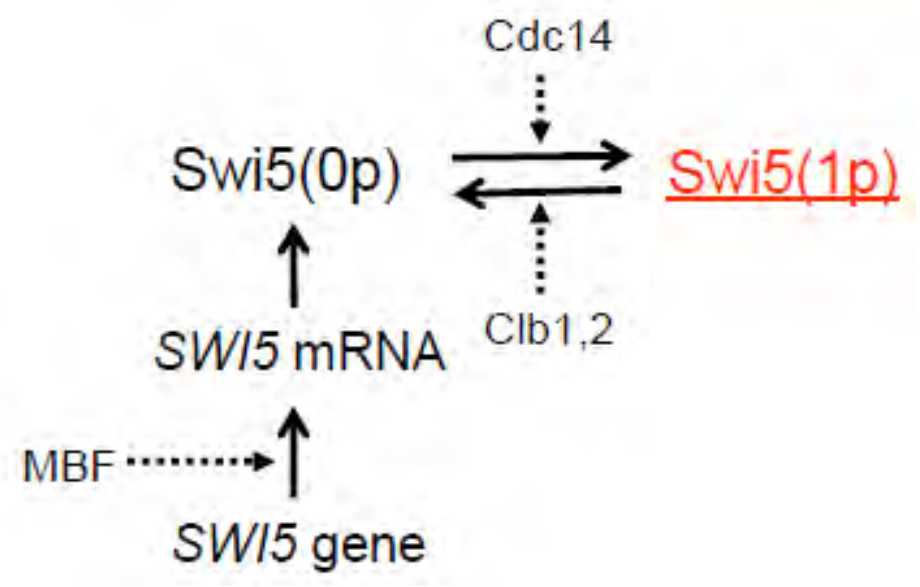

**References**

1. Mai B, Miles S, Breeden LL (2002) *Mol. Cell Biol.* 22: 430-441.
2. Simon I *et al.* (2001) *Cell* 106: 697-708.
3. MacKay VL, Mai B, Waters L, Breeden LL (2001) *Mol. Cell Biol.* 21: 4140-4148.
4. Levine K, Huang K, Cross FR (1996) *Mol. Cell Biol.* 16: 6794-6803.
5. Miller ME, Cross FR (2001) *Mol. Cell Biol.* 21: 6292-6311.
6. Yaglom J *et al.* (1995) *Mol. Cell Biol.* 15: 731-741.
7. Cross FR, Blake CM (1993) *Mol. Cell Biol.* 13: 3266-3271.
8. Tyers M, Tokiwa G, Nash R, Futcher B (1992) *EMBO J.* 11: 1773-1784.
9. Taylor IA *et al.* (2000) *Biochemistry* 39: 3943-3954.
10. Koch C, Schleiffer A, Ammerer G, Nasmyth K (1996) *Genes Dev.* 10: 129-141.
11. de Bruin RA *et al.* (2004) *Cell* 117: 887-898.
12. Amon A, Tyers M, Futcher B, Nasmyth K (1993) *Cell* 74: 993-1007.
13. Sidorova JM, Mikesell GE, Breeden LL (1995) *Mol. Biol. Cell* 6: 1641-1658.
14. Baetz K, Andrews B (1999) *Mol. Cell Biol.* 19: 6729-6741.
15. Loog M, Morgan DO (2005) *Nature* 434: 104-108.
16. Lee TI *et al.* (2002) *Science* 298: 799-804.
17. Stuart D, Wittenberg C (1994) *Mol. Cell Biol.* 14: 4788-4801.
18. Basco RD, Segal MD, Reed SI (1995) *Mol. Cell Biol.* 15: 5030-5042.
19. Kishi T, Yamao F (1998) *J. Cell Sci.* 111 (Pt 24): 3655-3661.
20. Skowyra D *et al.* (1997) *Cell* 91: 209-219.
21. Lanker S, Valdivieso MH, Wittenberg C (1996) *Science* 271: 1597-1601.
22. Edgington NP, Futcher B (2001) *J. Cell Sci.* 114: 4599-4611.
23. Moore SA (1988) *J. Biol. Chem.* 263: 9674-9681.
24. McBride HJ, Yu Y, Stillman DJ (1999) *J. Biol. Chem.* 274: 21029-21036.
25. Knapp D, Bhoite L, Stillman DJ, Nasmyth K (1996) *Mol. Cell Biol.* 16: 5701-5707.
26. Schwob E, Bohm T, Mendenhall MD, Nasmyth K (1994) *Cell* 79: 233-244.
27. Verma R *et al.* (1997) *Science* 278: 455-460.
28. Nash P *et al.* (2001) *Nature* 414: 514-521.
29. Bai C *et al.* (1996) *Cell* 86: 263-274.
30. Traverso EE *et al.* (2001) *J. Biol. Chem.* 276: 21924-21931.
31. Visintin R *et al.* (1998) *Mol. Cell* 2: 709-718.
32. Shirayama M, Toth A, Galova M, Nasmyth K (1999) *Nature* 402: 203-207.
33. Seufert W, Futcher B, Jentsch S (1995) *Nature* 373: 78-81.
34. Hood JK, Hwang WW, Silver PA (2001) *J. Cell Sci.* 114: 589-597.
35. Visintin R, Hwang ES, Amon A (1999) *Nature* 398: 818-823.

36. Loy CJ, Lydall D, Surana U (1999) *Mol. Cell Biol.* 19: 3312-3327.
37. Shirayama M, Zachariae W, Ciosk R, Nasmyth K (1998) *EMBO J.* 17: 1336-1349.
38. Prinz S, Hwang ES, Visintin R, Amon A (1998) *Curr. Biol.* 8: 750-760.
39. Jaspersen SL, Charles JF, Tinker-Kulberg RL, Morgan DO (1998) *Mol. Biol. Cell* 9: 2803-2817.
40. Irniger S, Piatti S, Michaelis C, Nasmyth K (1995) *Cell* 81: 269-278.
41. Althoefer H *et al.* (1995) *Mol. Cell Biol.* 15: 5917-5928.
42. Reynolds D *et al.* (2003) *Genes Dev.* 17: 1789-1802.
43. Darieva Z *et al.* (2003) *Curr. Biol.* 13: 1740-1745.
44. Jaquenoud M, van Drogen F, Peter M (2002) *EMBO J.* 21: 6515-6526.
45. Pan J, Chen RH (2004) *Genes Dev.* 18: 1439-1451.
46. Huang JN *et al.* (2001) *J. Cell Biol.* 154: 85-94.
47. Rudner AD, Hardwick KG, Murray AW (2000) *J. Cell Biol.* 149: 1361-1376.
48. Rudner AD, Murray AW (2000) *J. Cell Biol.* 149: 1377-1390.
49. Spellman PT *et al.* (1998) *Mol. Biol. Cell* 9: 3273-3297.
50. Cohen-Fix O, Peters JM, Kirschner MW, Koshland D (1996) *Genes Dev.* 10: 3081-3093.
51. Visintin R, Prinz S, Amon A (1997) *Science* 278: 460-463.
52. Shou W *et al.* (1999) *Cell* 97: 233-244.
53. Jaspersen SL, Charles JF, Morgan DO (1999) *Curr. Biol.* 9: 227-236.
54. Moll T *et al.* (1991) *Cell* 66: 743-758.
55. Zachariae W, Schwab M, Nasmyth K, Seufert W (1998) *Science* 282: 1721-1724.
56. Jans DA, Moll T, Nasmyth K, Jans P (1995) *J. Biol. Chem.* 270: 17064-17067.
57. Nasmyth K, Adolf G, Lydall D, Seddon A (1990) *Cell* 62: 631-647.

**Supporting Text2**

**Stable stochastic dynamics in yeast cell cycle**

**Yurie Okabe and Masaki Sasai**

The equations for the moments of states of genes, numbers of mRNA molecules, and numbers of protein molecules residing in each of chemical states are derived from the master equation. Those equations are approximated by truncating them at the 2nd order of cumulants and by neglecting the cross correlation between different molecular species. For a concrete example, equations for the moments related to the changes in the number of Clb5,6 are shown below. Equations for other 12 sets of proteins, mRNA and genes in the network can be derived in the same way.

Variables that appear in equations are the following: $\mu$ specifies the protein considered. In the present Supporting Text, $\mu =$Clb5,6. $\delta_{i,j}$ and $\delta_{R_\mu,j}$ are the Kronecker deltas. $i$ = 1, 2, ..., or 5 denotes the stage of cell cycle at time $t$ and $R_\mu$ = 1 or 2 represents the number of copies of the *clb*5,6 gene. $R_\mu$ increases from 1 to 2 by replication of the gene during stage1 and decreases from 2 to 1 by cytokinesis at $C_5$. The index $\alpha$ represents the gene state, $\alpha$ = 1 or 0 when $R_\mu$ = 1 and $\alpha$ is a pair of numbers, $\alpha$ = 11, 10, 01 or 00 when $R_\mu$ = 2. $D_\alpha^{\text{int}}(\mu,t)$ is the probability that $\mu$ th gene is in the state of α. $N_{m\alpha}^{\text{int}}(\mu,t)$ is the mean number of mRNA molecules of *clb*5,6 at time *t* when the *clb*5,6 gene is in the state α. $N_X^{\text{int}}(\mu,t)$ is the mean number of Clb5,6 protein molecules at the chemical state *X*. We write the mean square of the number of mRNA molecules as $M_{m\alpha}^{\text{int}}(\mu,t)$ and the mean square of the number of Clb5,6 protein molecules as $M_X^{\text{int}}(\mu,t)$. Variances are then calculated as $\sigma_{m\alpha}^{\text{int}}(\mu,t) = M_{m\alpha}^{\text{int}}(\mu,t) - \left(N_{m\alpha}^{\text{int}}(\mu,t)\right)^2$ and $\sigma_X^{\text{int}}(\mu,t) = M_X^{\text{int}}(\mu,t) - \left(N_X^{\text{int}}(\mu,t)\right)^2$.

Kinetic parameters in the equations are coefficients of rates of reactions: translation ($\eta$), protein-complex formation ($h_b$), ubiquitination ($h_u$), degradation of ubiquinated protein ($k_0$), degradation of unubiquinated protein ($k_1$), degradation of mRNA ($k_2$), dissociation of activator from DNA ($f$), binding of activator to DNA ($h_t$), synthesis of mRNA from a single copy of the gene in the off state ($g_0$), and synthesis of mRNA from a single copy of the gene in the on state ($g_1$).

Since equations contain the index *i* representing the stage in cell cycle at time *t* and $R_\mu$ of the number of copies of the $\mu$th gene, the equations have different nonzero terms depending on $i(t)$ and $R_\mu(t)$. Equations are not self-contained to determine $i(t)$ and $R_\mu$(t) but $i(t)$ and $R_\mu$(t) are changed by following the stochastic rules defined independently of the equations of moments. See the main text for the rules to change $i(t)$ and $R_\mu$(t). When *i* is changed at $C_1$, $C_2$, $C_3$, or $C_4$, $D_\alpha^{\text{int}}(\mu,t)$, $N_{m\alpha}^{\text{int}}(\mu,t)$, $N_X^{\text{int}}(\mu,t)$, $M_{m\alpha}^{\text{int}}(\mu,t)$, and $M_X^{\text{int}}(\mu,t)$ are handed continuously to the next stage. When $i$ is changed at $C_5$ (i.e., at cytokinesis) from $i$ = 4 at time $t$ to $i$ = 5 at time $t+\Delta t$, $D_{\alpha}^{\text{int}}(\mu,t)$ is determined to be $D_{\xi}^{\text{int}}(\mu,t+\Delta t) = D_{\xi 1}^{\text{int}}(\mu,t) + D_{\xi 0}^{\text{int}}(\mu,t)$ and $N_{m\alpha}^{\text{int}}(\mu,t)$ and $N_{X}^{\text{int}}(\mu,t)$ are stochastically reduced roughly half as described in the main text. $M_{m\alpha}^{\text{int}}(\mu,t)$ and $M_{X}^{\text{int}}(\mu,t)$ are handed to make $F_{m\alpha}^{\text{int}}(\mu,t)$ and $F_{X}^{\text{int}}(\mu,t)$ continuous at $C_5$. When $R_{\mu}$ is increased on replication of the $\mu$th gene from $R_{\mu} = 1$ at time $t$ to $R_{\mu} = 2$ at time $t+\Delta t$, $D_{\alpha}^{\text{int}}(\mu,t)$, $N_{m\alpha}^{\text{int}}(\mu,t)$ and $M_{m\alpha}^{\text{int}}(\mu,t)$ are determined as $D_{\xi\xi'}^{\text{int}}(\mu,t+\Delta t) = D_{\xi}^{\text{int}}(\mu,t) \cdot D_{\xi'}^{\text{int}}(\mu,t)$, $N_{m\xi\xi'}^{\text{int}}(\mu,t+\Delta t) = N_{m\xi}^{\text{int}}(\mu,t) + N_{m\xi'}^{\text{int}}(\mu,t)$, and $M_{m\xi\xi'}^{\text{int}}(\mu,t+\Delta t) = 2\left(M_{m\xi}^{\text{int}}(\mu,t) + N_{m\xi}^{\text{int}}(\mu,t) \cdot N_{m\xi'}^{\text{int}}(\mu,t)\right)$, which ensures continuity of $F_{m\alpha}^{\text{int}}(\mu,t)$. $N_{X}^{\text{int}}(\mu,t)$ and $M_{m\alpha}^{\text{int}}(\mu,t)$ are $N_{X}^{\text{int}}(\mu,t+\Delta t) = N_{X}^{\text{int}}(\mu,t)$ and $M_{X}^{\text{int}}(\mu,t+\Delta t) = M_{X}^{\text{int}}(\mu,t)$. Duration of each stage, timing of replication of each gene and the ratio of distribution of molecules at cytokinesis are fluctuating at every cycle as described in the main text, and thus these cycle-by-cycle variations are the origins of extrinsic fluctuations in the present model.

< Equations for the numbers of protein molecules >

$$\frac{d}{dt}N_{(1u)}^{\mathrm{int}}(\mu,t)=\eta\cdot\left(\delta_{R_\mu,1}\sum_{\alpha=1\text{ or }0}D_\alpha^{\mathrm{int}}(\mu,t)N_{m\alpha}^{\mathrm{int}}(\mu,t)+\delta_{R_\mu,2}\sum_{\alpha=11,\,10,\,01,\text{ or }00}D_\alpha^{\mathrm{int}}(\mu,t)N_{m\alpha}^{\mathrm{int}}(\mu,t)\right)$$
$$-\left\{k_1+h_b\sum_X N_X^{\mathrm{int}}(Sic1,t)+(\delta_{i,3}+\delta_{i,4}+\delta_{i,5})h_u\left(N_{(1u)(1p)}^{\mathrm{int}}(Cdc20,t)+N_{(0u)(1p)}^{\mathrm{int}}(Cdc20,t)\right)\right\}N_{(1u)}^{\mathrm{int}}(\mu,t)$$

$$\frac{d}{dt}N_{(0u)}^{\mathrm{int}}(\mu,t)=-\left(k_0+h_b\sum_X N_X^{\mathrm{int}}(Sic1,t)\right)N_{(0u)}^{\mathrm{int}}(\mu,t)+(\delta_{i,3}+\delta_{i,4}+\delta_{i,5})h_u\left(N_{(1u)(1p)}^{\mathrm{int}}(Cdc20,t)+N_{(0u)(1p)}^{\mathrm{int}}(Cdc20,t)\right)N_{(1u)}^{\mathrm{int}}(\mu,t)$$

$$\frac{d}{dt}M_{(1u)}^{\mathrm{int}}(\mu,t)=\eta\cdot\left(\delta_{R_\mu,1}\sum_{\alpha=1\text{ or }0}D_\alpha^{\mathrm{int}}(\mu,t)N_{m\alpha}^{\mathrm{int}}(\mu,t)+\delta_{R_\mu,2}\sum_{\alpha=11,\,10,\,01,\text{ or }00}D_\alpha^{\mathrm{int}}(\mu,t)N_{m\alpha}^{\mathrm{int}}(\mu,t)\right)\left(2N_{(1u)}^{\mathrm{int}}(\mu,t)+1\right)$$
$$-\left\{k_1+h_b\sum_X N_X^{\mathrm{int}}(Sic1,t)+(\delta_{i,3}+\delta_{i,4}+\delta_{i,5})h_u\left(N_{(1u)(1p)}^{\mathrm{int}}(Cdc20,t)+N_{(0u)(1p)}^{\mathrm{int}}(Cdc20,t)\right)\right\}\left(2M_{(1u)}^{\mathrm{int}}(\mu,t)-N_{(1u)}^{\mathrm{int}}(\mu,t)\right)$$

$$\frac{d}{dt}M_{(0u)}^{\mathrm{int}}(\mu,t)=-\left(k_0+h_b\sum_X N_X^{\mathrm{int}}(Sic1,t)\right)\left(2M_{(0u)}^{\mathrm{int}}(\mu,t)-N_{(0u)}^{\mathrm{int}}(\mu,t)\right)$$
$$+(\delta_{i,3}+\delta_{i,4}+\delta_{i,5})h_u\left(N_{(1u)(1p)}^{\mathrm{int}}(Cdc20,t)+N_{(0u)(1p)}^{\mathrm{int}}(Cdc20,t)\right)N_{(1u)}^{\mathrm{int}}(\mu,t)\left(2N_{(0u)}^{\mathrm{int}}(\mu,t)+1\right)$$

< Equations for the state of gene and the number of mRNA molecules <u>before replication of the $\mu$th gene</u> >

$$\frac{d}{dt}D_1^{\text{int}}(\mu,t)=\delta_{R_\mu,1}\left[-f\,D_1^{\text{int}}(\mu,t)+h_t\left(M_{(1p)(1p)}^{\text{int}}(MBF,t)-N_{(1p)(1p)}^{\text{int}}(MBF,t)\right)D_0^{\text{int}}(\mu,t)\right]$$

$$\frac{d}{dt}D_0^{\text{int}}(\mu,t)=\delta_{R_\mu,1}\left[f\,D_1^{\text{int}}(\mu,t)-h_t\left(M_{(1p)(1p)}^{\text{int}}(MBF,t)-N_{(1p)(1p)}^{\text{int}}(MBF,t)\right)D_0^{\text{int}}(\mu,t)\right]$$

$$\frac{d}{dt}D_1^{\text{int}}(\mu,t)N_{m1}^{\text{int}}(\mu,t)=\delta_{R_\mu,1}\left[D_1^{\text{int}}(\mu,t)\left\{g_1-(k_2+f)N_{m1}^{\text{int}}(\mu,t)\right\}+h_t\left(M_{(1p)(1p)}^{\text{int}}(MBF,t)-N_{(1p)(1p)}^{\text{int}}(MBF,t)\right)D_0^{\text{int}}(\mu,t)N_{m0}^{\text{int}}(\mu,t)\right]$$

$$\frac{d}{dt}D_0^{\text{int}}(\mu,t)N_{m0}^{\text{int}}(\mu,t)=\delta_{R_\mu,1}\left[D_0^{\text{int}}(\mu,t)\left\{g_0-\left(k_2+h_t\left(M_{(1p)(1p)}^{\text{int}}(MBF,t)-N_{(1p)(1p)}^{\text{int}}(MBF,t)\right)\right)N_{m0}^{\text{int}}(\mu,t)\right\}+f\,D_1^{\text{int}}(\mu,t)N_{m1}^{\text{int}}(\mu,t)\right]$$

$$\begin{aligned}\frac{d}{dt}D_1^{\text{int}}(\mu,t)M_{m1}^{\text{int}}(\mu,t)=\delta_{R_\mu,1}\Big[&D_1^{\text{int}}(\mu,t)\left\{g_1\left(2N_{m1}^{\text{int}}(\mu,t)+1\right)-k_2\left(2M_{m1}^{\text{int}}(\mu,t)-N_{m1}^{\text{int}}(\mu,t)\right)-f\,M_{m1}^{\text{int}}(\mu,t)\right\}\\&+h_t\left(M_{(1p)(1p)}^{\text{int}}(MBF,t)-N_{(1p)(1p)}^{\text{int}}(MBF,t)\right)D_0^{\text{int}}(\mu,t)M_{m0}^{\text{int}}(\mu,t)\Big]\end{aligned}$$

$$\begin{aligned}\frac{d}{dt}D_0^{\text{int}}(\mu,t)M_{m0}^{\text{int}}(\mu,t)=\delta_{R_\mu,1}\Big[&D_0^{\text{int}}(\mu,t)\left\{g_0\left(2N_{m0}^{\text{int}}(\mu,t)+1\right)-k_2\left(2M_{m0}^{\text{int}}(\mu,t)-N_{m0}^{\text{int}}(\mu,t)\right)-h_t\left(M_{(1p)(1p)}^{\text{int}}(MBF,t)-N_{(1p)(1p)}^{\text{int}}(MBF,t)\right)M_{m0}^{\text{int}}(\mu,t)\right\}\\&+f\,D_1^{\text{int}}(\mu,t)M_{m1}^{\text{int}}(\mu,t)\Big]\end{aligned}$$

< Equations for the state of gene and the number of mRNA molecules <u>after replication of the $\mu$th gene</u> >

$$\frac{d}{dt}D_{11}^{\text{int}}(\mu,t)=\delta_{R_\mu,2}\left[h_t\left(M_{(1p)(1p)}^{\text{int}}(MBF,t)-N_{(1p)(1p)}^{\text{int}}(MBF,t)\right)\left(D_{10}^{\text{int}}(\mu,t)+D_{01}^{\text{int}}(\mu,t)\right)-2f\,D_{11}^{\text{int}}(\mu,t)\right]$$

$$\frac{d}{dt}D_{10}^{\text{int}}(\mu,t)=\delta_{R_\mu,2}\left[h_t\left(M_{(1p)(1p)}^{\text{int}}(MBF,t)-N_{(1p)(1p)}^{\text{int}}(MBF,t)\right)\left(D_{00}^{\text{int}}(\mu,t)-D_{10}^{\text{int}}(\mu,t)\right)+f\left(D_{11}^{\text{int}}(\mu,t)-D_{10}^{\text{int}}(\mu,t)\right)\right]$$

$$\frac{d}{dt}D_{01}^{\text{int}}(\mu,t)=\delta_{R_\mu,2}\left[h_t\left(M_{(1p)(1p)}^{\text{int}}(MBF,t)-N_{(1p)(1p)}^{\text{int}}(MBF,t)\right)\left(D_{00}^{\text{int}}(\mu,t)-D_{01}^{\text{int}}(\mu,t)\right)+f\left(D_{11}^{\text{int}}(\mu,t)-D_{01}^{\text{int}}(\mu,t)\right)\right]$$

$$\frac{d}{dt}D_{00}^{\text{int}}(\mu,t)=\delta_{R_\mu,2}\left[-2h_t\left(M_{(1p)(1p)}^{\text{int}}(MBF,t)-N_{(1p)(1p)}^{\text{int}}(MBF,t)\right)D_{00}^{\text{int}}(\mu,t)+f\left(D_{10}^{\text{int}}(\mu,t)+D_{01}^{\text{int}}(\mu,t)\right)\right]$$

$$\begin{aligned}\frac{d}{dt}D_{11}^{\text{int}}(\mu,t)N_{m11}^{\text{int}}(\mu,t)=\delta_{R_\mu,2}\Big[&D_{11}^{\text{int}}(\mu,t)\left(2g_1-k_2N_{m11}^{\text{int}}(\mu,t)\right)-2f\,D_{11}^{\text{int}}(\mu,t)N_{m11}^{\text{int}}(\mu,t)\\&+h_t\left(M_{(1p)(1p)}^{\text{int}}(MBF,t)-N_{(1p)(1p)}^{\text{int}}(MBF,t)\right)\left(D_{10}^{\text{int}}(\mu,t)N_{m10}^{\text{int}}(\mu,t)+D_{01}^{\text{int}}(\mu,t)N_{m01}^{\text{int}}(\mu,t)\right)\Big]\end{aligned}$$

$$\begin{aligned}\frac{d}{dt}D_{10}^{\text{int}}(\mu,t)N_{m10}^{\text{int}}(\mu,t)=\delta_{R_\mu,2}\Big[&D_{10}^{\text{int}}(\mu,t)\left(g_1+g_0-k_2N_{m10}^{\text{int}}(\mu,t)\right)+f\left(D_{11}^{\text{int}}(\mu,t)N_{m11}^{\text{int}}(\mu,t)-D_{10}^{\text{int}}(\mu,t)N_{m10}^{\text{int}}(\mu,t)\right)\\&+h_t\left(M_{(1p)(1p)}^{\text{int}}(MBF,t)-N_{(1p)(1p)}^{\text{int}}(MBF,t)\right)\left(D_{00}^{\text{int}}(\mu,t)N_{m00}^{\text{int}}(\mu,t)-D_{10}^{\text{int}}(\mu,t)N_{m10}^{\text{int}}(\mu,t)\right)\Big]\end{aligned}$$

$$\begin{aligned}\frac{d}{dt}D_{01}^{\text{int}}(\mu,t)N_{m01}^{\text{int}}(\mu,t)=\delta_{R_\mu,2}\Big[&D_{01}^{\text{int}}(\mu,t)\left(g_1+g_0-k_2N_{m01}^{\text{int}}(\mu,t)\right)+f\left(D_{11}^{\text{int}}(\mu,t)N_{m11}^{\text{int}}(\mu,t)-D_{01}^{\text{int}}(\mu,t)N_{m01}^{\text{int}}(\mu,t)\right)\\&+h_t\left(M_{(1p)(1p)}^{\text{int}}(MBF,t)-N_{(1p)(1p)}^{\text{int}}(MBF,t)\right)\left(D_{00}^{\text{int}}(\mu,t)N_{m00}^{\text{int}}(\mu,t)-D_{01}^{\text{int}}(\mu,t)N_{m01}^{\text{int}}(\mu,t)\right)\Big]\end{aligned}$$

$$\frac{d}{dt}D_{00}^{\mathrm{int}}(\mu,t)N_{m00}^{\mathrm{int}}(\mu,t)=\delta_{R_\mu,2}\Big[D_{00}^{\mathrm{int}}(\mu,t)\big(2g_0-k_2N_{m00}^{\mathrm{int}}(\mu,t)\big)+f\big(D_{10}^{\mathrm{int}}(\mu,t)N_{m10}^{\mathrm{int}}(\mu,t)+D_{01}^{\mathrm{int}}(\mu,t)N_{m01}^{\mathrm{int}}(\mu,t)\big)$$
$$-2h_t\big(M_{(1p)(1p)}^{\mathrm{int}}(MBF,t)-N_{(1p)(1p)}^{\mathrm{int}}(MBF,t)\big)D_{00}^{\mathrm{int}}(\mu,t)N_{m00}^{\mathrm{int}}(\mu,t)\Big]$$

$$\frac{d}{dt}D_{11}^{\mathrm{int}}(\mu,t)M_{m11}^{\mathrm{int}}(\mu,t)=\delta_{R_\mu,2}\Big[D_{11}^{\mathrm{int}}(\mu,t)\big\{2g_1\big(2N_{m11}^{\mathrm{int}}(\mu,t)+1\big)-k_2\big(2M_{m11}^{\mathrm{int}}(\mu,t)-N_{m11}^{\mathrm{int}}(\mu,t)\big)\big\}-2f\,D_{11}^{\mathrm{int}}(\mu,t)M_{m11}^{\mathrm{int}}(\mu,t)$$
$$+h_t\big(M_{(1p)(1p)}^{\mathrm{int}}(MBF,t)-N_{(1p)(1p)}^{\mathrm{int}}(MBF,t)\big)\big(D_{10}^{\mathrm{int}}(\mu,t)M_{m10}^{\mathrm{int}}(\mu,t)+D_{01}^{\mathrm{int}}(\mu,t)M_{m01}^{\mathrm{int}}(\mu,t)\big)\Big]$$

$$\frac{d}{dt}D_{10}^{\mathrm{int}}(\mu,t)M_{m10}^{\mathrm{int}}(\mu,t)=\delta_{R_\mu,2}\Big[D_{10}^{\mathrm{int}}(\mu,t)\big\{(g_1+g_0)\big(2N_{m10}^{\mathrm{int}}(\mu,t)+1\big)-k_2\big(2M_{m10}^{\mathrm{int}}(\mu,t)-N_{m10}^{\mathrm{int}}(\mu,t)\big)\big\}+f\big(D_{11}^{\mathrm{int}}(\mu,t)M_{m11}^{\mathrm{int}}(\mu,t)-D_{10}^{\mathrm{int}}(\mu,t)M_{m10}^{\mathrm{int}}(\mu,t)\big)$$
$$+h_t\big(M_{(1p)(1p)}^{\mathrm{int}}(MBF,t)-N_{(1p)(1p)}^{\mathrm{int}}(MBF,t)\big)\big(D_{00}^{\mathrm{int}}(\mu,t)M_{m00}^{\mathrm{int}}(\mu,t)-D_{10}^{\mathrm{int}}(\mu,t)M_{m10}^{\mathrm{int}}(\mu,t)\big)\Big]$$

$$\frac{d}{dt}D_{01}^{\mathrm{int}}(\mu,t)M_{m01}^{\mathrm{int}}(\mu,t)=\delta_{R_\mu,2}\Big[D_{01}^{\mathrm{int}}(\mu,t)\big\{(g_1+g_0)\big(2N_{m01}^{\mathrm{int}}(\mu,t)+1\big)-k_2\big(2M_{m01}^{\mathrm{int}}(\mu,t)-N_{m01}^{\mathrm{int}}(\mu,t)\big)\big\}+f\big(D_{11}^{\mathrm{int}}(\mu,t)M_{m11}^{\mathrm{int}}(\mu,t)-D_{01}^{\mathrm{int}}(\mu,t)M_{m01}^{\mathrm{int}}(\mu,t)\big)$$
$$+h_t\big(M_{(1p)(1p)}^{\mathrm{int}}(MBF,t)-N_{(1p)(1p)}^{\mathrm{int}}(MBF,t)\big)\big(D_{00}^{\mathrm{int}}(\mu,t)M_{m00}^{\mathrm{int}}(\mu,t)-D_{01}^{\mathrm{int}}(\mu,t)M_{m01}^{\mathrm{int}}(\mu,t)\big)\Big]$$

$$\frac{d}{dt}D_{00}^{\mathrm{int}}(\mu,t)M_{m00}^{\mathrm{int}}(\mu,t)=\delta_{R_\mu,2}\Big[D_{00}^{\mathrm{int}}(\mu,t)\big\{2g_0\big(2N_{m00}^{\mathrm{int}}(\mu,t)+1\big)-k_2\big(2M_{m00}^{\mathrm{int}}(\mu,t)-N_{m00}^{\mathrm{int}}(\mu,t)\big)\big\}+f\big(D_{10}^{\mathrm{int}}(\mu,t)M_{m10}^{\mathrm{int}}(\mu,t)+D_{01}^{\mathrm{int}}(\mu,t)M_{m01}^{\mathrm{int}}(\mu,t)\big)$$
$$-2h_t\big(M_{(1p)(1p)}^{\mathrm{int}}(MBF,t)-N_{(1p)(1p)}^{\mathrm{int}}(MBF,t)\big)D_{00}^{\mathrm{int}}(\mu,t)M_{m00}^{\mathrm{int}}(\mu,t)\Big]$$

Supplementary Table 1. Standard, minimum, and maximum values of parameters

| Category | Parameter | STD ($min^{-1}$) | MIN ($min^{-1}$) | MAX ($min^{-1}$) |
|---|---|---|---|---|
| binding constant of activator to DNA | $h_t$ | 0.002 | $9 \times 10^{-4}$ | $5 \times 10^{-3}$ |
| dissociation constant of activator from DNA | $f$ | 0.08 | 0.05 | 0.16 |
| transcription rate constant in the ON state | $g_1$ | 2 | 1.7 | 4 |
| transcription rate constant in the OFF state | $g_0$ | 0.02 | $< 1 \times 10^{-7}$ | 0.9 |
| transcription rate constant in the constitutive expression ** | $g_2$ | 0.2 | 0.1 | 0.6 |
| degradation rate constant of mRNA ** | $k_2$ | ln2/5 | ln2/9 | ln2/4 |
| translation rate constant | $\eta$ | 0.2 | 0.15 | 0.3 |
| phosphorylation rate constant | $h_p$ | $1 \times 10^{-2}$ | $5 \times 10^{-4}$ | 0.2 |
| dephosphorylation rate constant | $h_p'$ | $2 \times 10^{-2}$ | $2 \times 10^{-3}$ | > 7 |
| ubiquitination rate constant | $h_u$ | $5 \times 10^{-3}$ | $< 1 \times 10^{-5}$ | > 4 |
| complex formation | $h_b$ | 0.1 | $< 1 \times 10^{-6}$ | > 10 |
| degradation rate constant of unubiquitinated proteins * | $k_1$ | ln2/120 | $< \ln2/10^4$ | ln2/40 |
| degradation rate constant of ubiquitinated proteins * | $k_0$ | ln2/5 | $< \ln2/10^4$ | $> \ln2/10^{-3}$ |
| exporting rate constant of Cdc14 | $h_r$ | ln2/160 | $\ln2/10^3$ | $> \ln2/10^{-2}$ |
| importing rate constant of Cdc14 | $h_r'$ | ln2/10 | $< \ln2/10^5$ | ln2/0.5 |

More than 300 reactions represented by the master equation are categorized into 15 types and a single parameter is assigned to each reaction type. Values in the third right column (STD) are used as the standard parameters. The standard limit cycle is robust when one of 15 parameters is varied in the range from the value of the second right column (MIN) to that of the first right column (MAX).

*Half-lives of ubiquitinated proteins are several minutes or less (see *Supporting Text*) and those of unubiquitinated proteins are >60 min (1-3).

**Transcription frequencies of *CLN1,2*, *CLN3*, *CLB1,2*, *CLB5,6*, *SIC1*, *CDC20*, *SWI5*, and *PDS1* are estimated to be ~$10^{-2}$ mRNAs/min for asynchronous cells and half-lives of these mRNAs are estimated to be 9-17 min (4). These half-lives should include time needed for the export process of mRNA to cytoplasm. Because mRNA can be translated into proteins immediately after transcription in the present model, we set half-life of mRNA 5 min which is shorter than the experimental results.

**References**

1. Lanker, S., Valdivieso, M. H. & Wittenberg, C. (1996) *Science* **271,** 1597-601.
2. Rudner, A. D., Hardwick, K. G. & Murray, A. W. (2000) *J. Cell Biol.* **149,** 1361-76.
3. Seufert, W., Futcher, B. & Jentsch, S. (1995) *Nature* **373,** 78-81.
4. Holstege, F. C., Jennings, E. G., Wyrick, J. J., Lee, T. I., Hengartner, C. J., Green, M. R., Golub, T. R., Lander, E. S. & Young, R. A. (1998) *Cell* **95,** 717-28.

Supplementary Table 2. Peaks of the mRNA levels

| Gene | Experimental results (Ref.1) | Experimental results (Ref.2) | Model results |
|---|---|---|---|
| *CLN3* | M | M/G1 | M/G1 |
| *CLN1,2* | late G1 | late G1 | S/G2/M |
| *SIC1* | early G1 | M/G1 | M/G1 |
| *CLB5,6* | late G1 | late G1 | S/G2/M |
| *NDD1* | - | S | S/G2 |
| *CLB1,2* | M | M | S/G2/M |
| *CDC20* | M | - | M |
| *PDS1* | S | - | S/G2/M |
| *SWI5* | M | M | M |

Peaks of the mRNA levels observed in experiments and those in the present model are compared. mRNA levels are measured by concentration in experiments and by numbers in the present model.

References

1. Cho RJ, Campbell MJ, Winzeler EA, Steinmetz L, Conway A, Wodicka L, Wolfsberg TG, Gabrielian AE, Landsman D, Lockhart DJ, and Davis RW (1998) *Mol. Cell* 2: 65-73.
2. Breeden LL (2003) *Curr.Biol.* 13: R31-38.

Supplementary Table 3. Temporal patterns of the protein levels

| Protein | Experimental observations (concentration) | Model results (number)* |
|---|---|---|
| Cln3 | The Cln3 level exhibits moderate periodicities in amplitude [1]. | The Cln3 level increases about two fold in G1 phase and decreases during S phase. |
| Cln1,2 | The Cln1,2 level and the Cln1,2/Cdc28 kinase activity are maximal at START [1]. | The Cln1,2 level is maximal in M phase. |
| Sic1 | Sic1 disappears at the G1/S transition and does not reappear until cell division [2]. | The Sic1 level is low at the metaphase/anaphase boundary, peaks in late G1, and decreases during S phase. |
| Clb5,6 | Clb5/Cdc28 kinase activity peaks around S phase [2].<br>Clb5 localizes in the nucleus until shortly before the metaphase/anaphase transition, and then it disappears during anaphase [3]. | The Clb5,6 and Clb5,6/Cdc28 levels increase during S phase, peak around the metaphase/anaphase boundary, and start declining during anaphase. |
| Ndd1 | The phosphorylated Ndd1 level increases in the G1/S transition, remains high in G2 phase, and declines in M phase [4]. | The phosphorylated Ndd1 level increases during S and G2 phases, peaks at the metaphase/anaphase boundary, and decreases during anaphase. |
| Clb1,2 | The Clb2 level peaks in M phase [5] and starts declining during anaphase B [6]. | The Clb1,2 level increases during S phase, peaks at the metaphase/anaphase boundary, and decreases during anaphase. |
| Cdc20 | The Cdc20 level is low in S phase, peaks in M phase, and declines at the M/G1 boundary [5]. | The Cdc20 level is low in S phase, increases until the anaphase/telophase boundary, and then decreases. |
| Pds1 | Pds1 disappears shortly before the onset of anaphase [6, 7]. | The Pds1 level is low in G1 phase, increases during S and G2 phases, and then starts declining in metaphase. |
| Swi5 | Prior to anaphase, Swi5 is phosphorylated and localizes in cytoplasm, but it is dephosphorylated and translocated to nucleus around the anaphase/telophase boundary [8, 9].<br>Most of Swi5 located in the nucleus is degraded by the time of cell separation [9]. | The total Swi5 level peaks in M phase. The phosphorylated Swi5 level, which is assumed to locate in nucleus, sharply peaks at around the anaphase/telophase boundary. |

*The protein levels are observed in concentration in experiments but in molecular numbers in the model.

References
1. Cross FR, and Blake CM (1993) *Mol. Cell Biol.* 13: 3266-3271.
2. Schwob E, Bohm T, Mendenhall MD, and Nasmyth K (1994) *Cell* 79: 233-244.
3. Shirayama M, Toth A, Galova M, and Nasmyth K (1999) *Nature* 402: 203-207.
4. Darieva Z, Pic-Taylor A, Boros J, Spanos A, Geymonat M, Reece RJ, Sedgwick SG, Sharrocks AD, and Morgan BA (2003) *Curr. Biol.* 13: 1740-1745.
5. Prinz S, Hwang ES, Visintin R, and Amon A (1998) *Curr. Biol.* 8: 750-760.
6. Shirayama M, Zachariae W, Ciosk R, and Nasmyth K (1998) *EMBO J.* 17: 336-1349.
7. Jaspersen SL, Charles JF, Tinker-Kulberg RL, and Morgn DO (1998) *Mol. Biol. Cell* 9: 2803-2817.
8. Moll T, Tebb G., Surana U, Robitsch H, and Nasmyth K (1991) *Cell* 66: 743-758.
9. Nasmyth K, Adolf G, Lydall D, and Seddon A (1990) *Cell* 62: 631-647.